\documentclass[traditabstract]{aa}
\usepackage{txfonts}
\usepackage{graphicx}
\usepackage{color}
\usepackage{placeins}

\begin{document}

\title{Metal-poor stars towards the Galactic bulge -- \\a population potpourri\thanks{This paper includes data gathered with the 6.5 meter Magellan Telescopes located at Las Campanas Observatory, Chile.
Tables 2 and 3 are only available in electronic form
at the CDS via anonymous ftp to cdsarc.u-strasbg.fr (130.79.128.5)
or via http://cdsweb.u-strasbg.fr/cgi-bin/qcat?J/A+A/}}

\author{
Andreas Koch\inst{1},   
Andrew McWilliam\inst{2},  
George W. Preston\inst{2},
\and 
Ian B. Thompson\inst{2}
}
\authorrunning{A. Koch et al.}
\titlerunning{Metal-poor stars towards the bulge}
\offprints{A. Koch; \email{akoch@lsw.uni-heidelberg.de; \\andy,gwp,ian@obs.carnegiescience.edu}}
\institute{
Landessternwarte, Zentrum f\"ur Astronomie der Universit\"at Heidelberg, K\"onigstuhl 12, 69117 Heidelberg, Germany
\and
Carnegie Observatories, 813 Santa Barbara St., Pasadena, CA 91101, USA}
\date{Received: 21 September 2015 / Accepted: 30 October 2015}
\abstract{
We present a comprehensive chemical abundance analysis of five red giants 
and two horizontal branch (HB) stars 
towards the southern edge of the Galactic bulge, at ($l$, $b$)$\sim$(0\degr,$-11$\degr). 
Based on high-resolution spectroscopy obtained with the Magellan/MIKE spectrograph, we 
derived  up to 23 chemical element abundances and identify a mixed bag of stars, representing
various populations in the central regions of the Galaxy. 
Although cosmological simulations predict that  the inner Galaxy was host to the first stars in the Universe, we see no 
chemical evidence of 
the ensuing massive supernova explosions: all of our targets exhibit halo-like, solar  [Sc/Fe] ratios, which is in contrast to the low values 
predicted from Population III nucleosynthesis. 
One of the targets is a CEMP-$s$ star at [Fe/H]=$-2.52$ dex, and another target is a moderately metal-poor ([Fe/H]=$-1.53$ dex)
CH star with strong enrichment in $s$-process elements (e.g., [Ba/Fe]=1.35).  
{These individuals provide the first contenders of these classes of stars towards the bulge}. 
Four of the carbon-normal stars exhibit abundance patterns reminiscent of halo star across a metallicity range spanning $-2.0$ to $-2.6$ dex, 
i.e., enhanced $\alpha$-elements and solar Fe-peak and neutron-capture elements, and the remaining one is a regular metal-rich bulge giant.
The position, distance, and radial velocity of one of the metal-poor HB stars coincides with simulations of the old trailing arm of the disrupted Sagittarius dwarf galaxy. 
While their highly uncertain proper motions prohibit a clear kinematic separation,  
the stars' chemical abundances and distances suggest that these metal-poor candidates, albeit located towards the bulge,  
are not of the bulge, but rather inner halo stars on orbits that make them pass through the central regions. 
Thus, we caution similar claims of detections of metal-poor stars as true habitants of the bulge. 
}
\maketitle
%
%
%
%
\section{Introduction}
The first stars in the Universe were very massive and perished after very short lifetimes of a few tens of Myr (e.g., Bromm \& Larson 2004); 
however, it is still possible to see their nucleosynthetic  imprints on the next generation of long-lived stars that formed out of their ejecta (e.g., Caffau et al. 2013). 
{ The Galactic halo hosts a substantial fraction of known metal-poor stars:}
its metallicity distribution function (MDF) peaks at $-1.5$ dex and comprises a broad tail of metal-poor stars down to $-7$ dex (Beers \& Christlieb 2005; 
Schoerck et al. 2009; Keller et al. 2015).
Conversely, the Galactic bulge has been identified as an archetypical old (Clarkson et al. 2008), yet metal-rich population (e.g., McWilliam \& Rich 1994),
which indicates a rapid formation at early times. 

Intriguingly, chemodynamical models of of the chemical evolutionary history of the Milky Way's (MW) stellar halo 
predict that the oldest and most metal-poor stars in our Galaxy are most common in the bulge. 
The simulations of Diemand et al. (2005) already indicate that 30--60\% of the first stars would reside within the inner 3 kpc.
Similarly, Tumlinson (2010) suggests that metal-poor stars below $-3.5$ dex, formed at redshifts $z\ga15$, would be found 
most frequently in the bulge, owing to the inside-out growth of dark-matter-dominated large-scale structures.  
It is important to note, however, that the present-day, central location of these stars then does not reflect their birthplace, which is 
commonly paraphrased as ``in the bulge, not of the bulge''.

Within the hierarchical framework of structure formation, the Galactic stellar halo formed via the accretion of dwarf galaxy-like fragments. 
This was challenged by an apparent mismatch of the  chemical abundances of halo stars and the supposed building blocks (the dwarf spheroidal galaxies; e.g., Shetrone et al. 2001),  
except for the most metal-poor component, and when compared to the lowest-mass systems (e.g., Koch \& Rich 2014). 
The case is less clear in the bulge of the MW: an accretion origin of the bulge would leave kinematic traces in phase space, none of which has yet been conclusively detected 
in kinematic surveys (Howard et al. 2008; Kunder et al. 2012, 2014; but cf. Nidever et al. 2012).

 By now, metal-poor stars have been detected towards the bulge, but
 barely overlapping with the class of extremely metal-poor (EMP) stars, i.e., below $-3$ dex.
Moreover,  this very metal-poor component is only sparsely sampled, with only 27 spectroscopically confirmed stars { between $-2$ dex
 and the most metal-poor bulge candidate at $-$3.02 dex} 
(Ness et al. 2013a; Garc\'\i a P\'erez et al. 2013; Howes et al. 2014; Schlaufman \& Casey 2015; Casey \& Schlaufman 2015).
%
%
Nonetheless, the bulge offers a wealth of stars with chemical peculiarities of one kind or the other, which 
provide important insight into its chemical evolutionary history and covering, amongst others,  Li-rich giants,  $r$-process enhanced stars, and barium stars,  
(Johnson et al. 2012, 2013b; Lebzelter et al. 2013), albeit at higher metallicities above $\sim -1.8$ dex.
Thus, a comprehensive chemical identification of the purported very metal-poor bulge is still pending
and the question as to which component they, chemically speaking, resemble, remains open. 
In this context, Ness et al. (2013a) anatomised the metallicity distribution of the bulge over a broad field of view, suggesting up to five  
 components. While strongly peaked at supersolar metallicities, this MDF also revealed a (low-number) overlap with the inner Galactic halo and the metal-weak thick disk.
Finally, Casey \& Schlaufman (2015) found the first metal-poor bulge candidates down to [Fe/H]=$-$3 dex; the detailed chemical composition of their three stars is 
compatible with the general
MW halo, 
with the exception of low Sc/Fe abundance ratios, indicative of enrichment by Population III ejecta.
 
Finding these metal-poor stars and ascertaining their membership with the bulge is not only of interest for the sake of 
reconstructing the chemical evolution of the bulge, but also has important implications for studying the 
morphology of the bulge. 
Number counts of the red clump, which led to the discoveries of, for instance, bar, boxy and X-shaped substructures (e.g., McWilliam \& Zoccali 2010; 
Wegg et al. 2015) 
employed certain cuts in colour-magnitude space. 
However, a prominent population of metal-poor, thus bluer, stars can be missed by  fixed colour criteria, which  
could lead to  biased, reconstructed spatial structures. 
 
 Here, we present a long-standing  campaign { and its results} to systematically look for metal-poor bulge stars .
 This paper is organised as follows: In \S2 we lay out our target selection and data processing; the abundance analyses are described and discussed out in detail in 
\S\S3,4. In \S5 we single out two stars with peculiar abundance distributions and 
in \S6 we address the question of their membership with the bulge before summarizing our findings in \S7.
\section{Target selection, observations, and data reduction}
\subsection{Identification of metal-poor candidates}
The stars studied here were identified in a search for  EMP 
stars in the Galactic bulge (Preston et al. unpublished), near $b=-10\degr$, 
employing the 2.5-m du Pont and 1-m Swope telescopes at Las Campanas Observatory.

EMP candidate stars were initially identified by weak
absorption from the Ca~K line, at 3933~\AA , using narrow-band imaging.
Line absorption was estimated from the flux transmitted by a 20~\AA\ wide
Ca~K filter compared to the flux in a 200~\AA\ wide continuum filter, both 
centred at 3933\AA . BVI images were also obtained to calibrate
the temperature-sensitivity of the resulting CaK index.
The CaK Index and B$-$V colours of the bulge stars were then used to obtain 
initial photometric metallicities, which we calibrated against similar photometry of
known EMP standard stars.

Not all bulge stars with a  low CaK index turned out to be metal poor.  For instance, sometimes the 
Ca~K photometry suffered from spurious systematic effects, such as a difficult to 
resolve, direct cosmic ray hit on the stellar image, or a diffraction spike from a 
nearby bright star.  M stars, with strong molecular bands, have weak CaK indices 
because heavy TiO blanketing drowns  out the Ca~K absorption.
Therefore, in order to eliminate false-positive detections, 
low-resolution (R$\sim$2,000), multi-object, confirmatory spectra of the photometric EMP
candidates were obtained using the duPont Wide-Field-CCD (WFCCD) grism spectrograph.
From these spectra, the measured equivalent width (EW) of the Ca~K line was used, in combination
with B$-$V colour, to estimate metallicity.

We note that none of our targets overlaps with the moderately to very metal-poor samples of Johnson et al. (2012, 2013a, 2014), Garc\'\i a P\'erez, et al. (2013),  Howes et al. (2014), or Schlaufman \& Casey (2015).
\subsection{High-resolution MIKE spectropscopy}
Observations of seven EMP candidates presented here were taken spread over six nights in July 2007 with a median seeing of 0.95$\arcsec$,
while individual exposures reached as high as 2$\arcsec$ and notably better conditions ($\sim0.6\arcsec$) during several nights.
Our chosen set-up included a 0.5$\arcsec$ slit, 2$\times$1 binning in spectral and spatial dimensions and resulted in a resolving power of 
$R\sim45,000$.  An observing log is given in Table~1.
\begin{table*}[!btp]
\caption{Observing log}
\centering
\begin{tabular}{ccccccc}
\hline\hline       
Star\tablefootmark{a} & IAU name & $\alpha$ (J2000.0) &$\delta$ (J2000.0) &Date &Exp. time [s] &S/N\tablefootmark{c} [pixel$^{-1}$] \\
\hline                  
 872-13971  & J183256.52-344529.3 & 18:32:56.52 & $-$34:45:29.28 & Jul-14-2007      &  3600 & 50/35/28 \\ 
 899-14135  & J183338.54-342940.9 & 18:33:38.54 & $-$34:29:40.91 & Jul-08,09-2007 & 11700 & 40/25/12 \\
 873-17221  & J183426.75-345554.4 & 18:34:26.75 & $-$34:55:54.39 & Jul-08-2007      &  4800 & 50/30/20 \\
 874-24995  & J183558.75-345839.1 & 18:35:58.75 & $-$34:58:39.14 & Jul-10-2007      &  4800 & 50/35/25 \\
 941-27793  & J183113.29-335148.3 & 18:31:13.29 & $-$33:51:48.25 & Jul-12-2007      &  6600 & 60/40/28 \\
 958-10464  & J183003.87-333423.6 & 18:30:03.87 & $-$33:34:23.57 & Jul-10,11-2007 &  7060 & 50/40/30 \\ 
 896-37860  & J183025.73-343130.0 & 18:30:25.73 & $-$34:31:30.01 & Jul-14-2007      &  3600 & 40/35/25 \\
\hline                  
\end{tabular}
\tablefoot{
\tablefoottext{a}{ID consisting of plate number and entry number in our photometric catalogue. Throughout the paper, we { use the second part of this ID as 
unique identifier of the stars for brevity.}}
\tablefoottext{b}{Given at 6500/4500/4000~\AA.}}
\end{table*}

The data were processed in a standard manner with the pipeline reduction package of Kelson (2000; 2003), which accounts for   
flat field division, order tracing from quartz lamp flats, and 
wavelength calibration using built-in Th-Ar lamp exposures that were taken immediately after each science exposure. 
Continuum-normalisation was performed by dividing the extracted spectra by a high-order polynomial.  
The final spectra reach signal-to-noise (S/N) ratios of 50 per pixel in the peak of the order containing H$\alpha$, declining towards $\sim$25 
in the bluest orders at 4000\AA~(see Table~1). 

Radial velocities were measured from a cross correlation of the heliocentrically corrected spectra 
against a typical red giant star template, yielding a precision of typically 0.5 km\,s$^{-1}$. 
\section{Abundance analysis}
Our chemical element abundance analysis builds on a standard EW absolute, gf, abundance analysis, which closely follows the procedures of our previous works
(e.g., Koch \& McWilliam et al. 2014).
In all following steps, the  2013 version of the stellar abundance code MOOG (Sneden 1973) was used.  
The line list is the same used in Koch \& McWilliam (2014; and references therein) for the similarly metal-poor globular cluster NGC 5897 ([Fe/H]=$-2.04$ dex), 
complemented by  transitions listed in  McWilliam
et al. (1995) and Sadakane et al. (2004). For vanadium, we employed the most recent log\,$gf$ measurements of Lawler et al. (2014) and Wood (2014). 
One of the targets turned out to be a metal-rich star so that most features used in the metal-poor spectra are too strong and saturated to give meaningful 
abundances. For this object, we resorted to the line list of Ruchti et al. (2015), which was optimised for studies of the metal-rich Galactic components. 
Some heavy elements  (e.g., La, Nd, Eu) have their dominant transitions in the blue spectral range, which is affected by heavy blending (e.g., Hansen et al. 2015) and we did not 
include these lines in following analyis of the metal-rich star.
Finally, we note that in all cases, but in particular for the CEMP-$s$ star 27793, we assured that all used lines were located in spectral regions unaffected by molecular
absorption. 

We included the effects of hyperfine splitting for Sc, Mn, Co, Sr, Ba, and Eu using the information given in McWilliam et al. (1995) and Mashonkina \& Christlieb (2014). 
Here we note its strong effect on Mn, where the hyperfine corrections can reach as high as 0.7 dex. 
Individual EWs were measured by fitting a Gaussian profile to the absorption lines 
using IRAF's {\em splot}, unless otherwise noted, where we employed spectral synthesis. 
The resulting line list  is detailed  in  Table~2 and in Table~3 for the metal-rich star, while individual elements and transitions are further discussed in Sect.~5.  
\begin{table*}[htbp]
\caption{Line list for the metal-poor sample}
\centering          
\begin{tabular}{cccrrrrrrr}
\hline\hline       
& {$\lambda$} &  {E.P.} &  {} & \multicolumn{6}{c}{EW [m\AA]} \\
\cline{5-10}
\raisebox{1.5ex}[-1.5ex]{Element} &  {[\AA]} &  {[eV]}  &\raisebox{1.5ex}[-1.5ex]{log\,$gf$} 
&  {13971} &  {17221} &  {24995} &  {27993} &  {10464} &  {37860} \\
\hline
Li I & 6707.70 &  0.00 &   0.174 &    18 &    15 & \dots & \dots & \dots & \dots \\
Na I & 5889.95 & 0.00 &   0.110 &   181 &   199 &   180 &   193 &   283 &   223 \\
Na I & 5895.92 & 0.00 &   0.190 &   159 &   167 &   162 &   173 &   252 &   211 \\ 
Na I & 8183.26 &  2.10 &   0.230 &    36 &    22 &    28 &    41 &   117 &    76 \\
Na I & 8194.79 &  2.10 &   0.470 &    54 &    22 &    53 &    88 &   153 &    88 \\
Mg I & 4571.10 &  0.00 &  $-$5.623 &    77 &    60 &    82 &    49 &    88 &    70 \\
Mg I & 4702.99 &  4.35 &  $-$0.440 &    87 &    77 &    87 &   111 &   149 &   110 \\
Mg I & 5528.42 &  4.35 &  $-$0.481 &    93 &    80 &    92 &   100 &   153 &   124 \\
Mg I & 5711.09 &  4.33 &  $-$1.728 & \dots &    19 &    22 &    11 &    57 &    24 \\
\hline
\hline
 \end{tabular}
\tablefoot{Table~2 is available in its entirety in electronic form via the CDS.}
\end{table*}
\begin{table}[htb]
\caption{Line list for the metal-rich giant 14135}
\centering          
\begin{tabular}{cccrr}
\hline\hline       
Element & {$\lambda$ [\AA] } &  {E.P. [eV]} &  {log\,$gf$} & EW [m\AA] \\
\hline
Na I & 5688.20 &  2.10 &  $-$0.404 &   168  \\
Na I & 6154.23 &  2.10 &  $-$1.547 &   111  \\
Na I & 6160.75 &  2.10 &  $-$1.246 &   130  \\
Mg I & 5711.09 &  4.35 &  $-$1.724 &   148  \\
Mg I & 6318.72 &  5.11 &  $-$2.103 &    83  \\
Mg I & 6319.24 &  5.11 &  $-$2.324 &    57  \\
Mg I & 6319.49 &  5.11 &  $-$2.803 &    30  \\
Mg I & 7387.69 &  5.75 &  $-$1.000 &   106  \\
Mg I & 7691.55 &  5.75 &  $-$0.783 &   124  \\
\hline
\hline
 \end{tabular}
\tablefoot{Table~3 is available in its entirety in electronic form via the CDS.}
\end{table}

Model atmospheres were interpolated from the ATLAS grid of Kurucz\footnote{\tt http://kurucz.harvard.edu/grids.html}, 
one-dimensional 72-layer, plane-parallel, line-blanketed models  
without convective overshoot, assuming local thermodynamic equilibrium (LTE) 
for all species. 
The models incorporate 
the $\alpha$-enhanced opacity distribution functions, 
AODFNEW (Castelli \& Kurucz 2003), since we anticipate the metal-poor bulge stars to show elevated  [$\alpha$/Fe] ratios, as is seen in 
comparably metal-poor halo field stars.
This is indeed in  accord with our finding of $\alpha$-enhancements in all of the target stars  (Sect.~5.3).  
The effect of using Solar-scaled distributions instead is negligible, as quantified  in Sect.~4.1 

Initial estimates for the effective temperatures were obtained from the targets' photometry and by using the colour-temperature relations of Alonso et al. (1999). 
To this end, we used the stellar B$-$V colours from Preston et al. (unpublished)
and J, H, and K-band magnitudes from the 2 Micron All Sky Survey (2MASS; Skrutskie et al. 2006),  which we transformed into the 
photometric system required by the Alonso calibrations, following Alonso et al. (1998) and Cutri (2003).
This photometry was dereddened with the Schlegel et al. (1998) maps and applying the reddening law of Winkler (1997). 
The average values for T$_{\rm eff}$ from four colour indices are given in Table~4. The typical uncertainty, as quantified by the standard deviation, is $\sim$130 K but always better than 200 K. 
\begin{table*}[htb]
\caption{Properties of the target stars}
\centering          
\begin{tabular}{cccccccccc}
\hline\hline       
 {} &   {$V$} &  {$(B-V)$} & {$(V-K)$} & \multicolumn{2}{c}{T$_{\rm eff}$ [K]}  &  {} &  {$\xi$} &  {} &  {}\\
 \cline{5-6}
\raisebox{1.5ex}[-1.5ex]{Star} &  {[mag]} &    {[mag]} & {[mag]}  &  {(phot.)\tablefootmark{a}} &  {(spec.)} & \raisebox{1.5ex}[-1.5ex]{log\,$g$}  &  {[km\,s$^{-1}$]} & \raisebox{1.5ex}[-1.5ex]{[Fe/H]} & \raisebox{1.5ex}[-1.5ex]{Type\tablefootmark{b}}\\
\hline
13971      	       & 14.92 & 0.806 &  2.48  & 4900$\pm$115		& 4900 & 1.83 & 1.64 & $-2.31$ & RGB \\ 
14135                  & 15.64 & 1.240  & 2.86  & 4485$\pm$55\phantom{0} & 4650 & 2.86 & 1.50 & $-0.01$ & RGB \\
17221      	       & 15.03 & 0.846 & 2.44  & 4866$\pm$144		& 4800 & 1.45 & 1.83 & $-2.66$ & RGB \\
24995      	       & 14.58 & 0.900 & 2.46  & 4921$\pm$186		& 4800 & 1.58 & 1.92 & $-2.50$ & RGB \\
27793 & 15.55 & 1.030 &  2.51  & 4906$\pm$187		& 4975 & 2.10 & 1.55 & $-2.52$ & RGB, CEMP-$s$ \\
10464 & 15.27 & 0.750 &  2.14  & 5500$\pm$108		& 5400 & 1.70 & 2.64 & $-1.53$ & HB, CH-star \\ 
37860     	       & 15.29 & 0.733 &  2.17  & 5296$\pm$70\phantom{0} & 5150 & 1.40 & 2.15 & $-2.07$ & HB \\
 \hline
 \hline
 \end{tabular}
\tablefoot{
\tablefoottext{a}{Average of T(V$-$K), T(J$-$K), T(J$-$H), and T(B$-$V). The error is the standard deviation of all four measures.}
\tablefoottext{b}{``RGB'' = red giant branch star; ``HB'' = Horizontal branch star.}
}
\end{table*}

Initial surface gravities, log\,$g$, were based on the canonical stellar structure equations, where we adopted the photometric  T$_{\rm eff}$ and the stars' 
V-band magnitude. The stellar mass was adopted as 0.8 M$_{\odot}$, typical of metal-poor red giants in old stellar populations. As the targets' distance
we adopted 8.5 kpc, assuming they are indeed located in the bulge (e.g., Reid et al. 2014).  
The CaK index was used to adopt the initial model atmosphere metallicities.

These stellar parameters were then refined by using an ensemble of Fe\,{\sc i} lines. First,  T$_{\rm eff}$ was set via excitation equilibrium, where we 
rejected weak lines with a reduced width RW=$\log ({\rm EW}/\lambda)$$<-5.5$ and strong, saturated lines above RW$>-$4.5. 
Thus, flat trends in the plot of abundance vs. excitation potential could be typically achieved within 50 K, yielding differences in the high- vs. low-excitation lines
of less than the intrinsic dispersion.  

Similarly, the microturbulence, $\xi$,  was set by removing any trend of abundance  with EW for this subset of lines, which yielded $\xi$ precise to $\sim$0.1 km\,s$^{-1}$.  
Finally, we determined spectroscopic surface gravities through ionisation equilibrium of neutral and ionised Fe lines. 
Typically, a variation of log\,$g$ by 0.15 dex yielded an ionisation imbalance less than the combined statistical error of the respective Fe\,{\sc i} and  Fe\,{\sc ii} abundances and we 
adopted this value as our uncertainty on the surface  gravity.  
These above procedures were iterated towards convergence of all parameters. 
All fiducial, spectroscopic, stellar parameters that we  use for the remainder of our analyses can be found in Table~4.  
We note a very good agreement of the photometric and spectroscopic temperatures, where the photometric values 
are warmer by 61 K on average with a 1$\sigma$ dispersion of 81 K. 

Two of the stars stars have lower gravities   than expected if they were on the red giant branch (RGB) or subgiant branch, given their warmer T$_{\rm eff}$.
We associate these with the horizontal branch (HB) and label them in Table~4. 
Our adopted temperatures for these two HB stars put them outside the T$_{\rm eff}$ range of the instability strip (e.g., see
Marconi et al. 2015).  Their spectra show no evidence of variability in terms of line asymmetries or other activity indicators (for instance within the Balmer lines), 
nor any radial velocity variations  between individual exposures, which were taken over the course of two nights (star 10464) and 
within one hour (object 37860). Thus we conclude that these stars are unlikely to be RR Lyrae or other variables. 
\section{Abundance results}
All of our following abundance results, presented in Tables 5,6 are relative to the Solar, photospheric scale of 
Asplund et al. (2009).
The subsequent Figures~1--5 show LTE abundances for the sake of comparison with literature and we discuss NLTE effects in the individual subsections.
\begin{table*}[htb]
\caption{Abundance results for the metal-poor stars with regular patterns.}             
\centering          
\begin{tabular}{crcrcrcrcrcrcrcr}     
\hline\hline       
 {} & {}  &  {13791}  & {}  & {}  &  {} &  {17221} & {}  &  {} &  {} 
&  {24995}  & {}  &  {} & {}  &  {37860}  &  {}   \\
\cline{2-4}\cline{6-8}\cline{10-12}\cline{14-16}
\raisebox{1.5ex}[-1.5ex]{Element\tablefootmark{a}} &  {[X/Fe]}  &  {$\sigma$} &  {N}  &  {} &  
 {[X/Fe]} &  {$\sigma$} &  {N}  &  {} &   {[X/Fe]} &  {$\sigma$} &  {N}  & {}  &  {[X/Fe]}&   {$\sigma$} &  {N} \\
 \hline
Fe\,{\sc i}        	       & $-2.31$ & 0.21 & 90 &  & $-2.66$ & 0.15 & 72 &  & $-2.49$ & 0.14 & 79 & & $-2.07$ & 0.22 & 87 \\ 
Fe\,{\sc ii}       	       & $-2.32$ & 0.13 &  5 &  &  $-2.66$ & 0.18 &   4 &  &  $-2.48$ & 0.12 & 4 & & $-2.06$ & 0.17 & 9 \\ 
A(Li)\rlap{$^{\rm LTE}$} & 0.90 &  \ldots & 1 &  & 0.69 &  \ldots & 1 &  &   \ldots &   \ldots & 0 & &  \ldots &  \ldots & 0 \\
A(Li)\rlap{$^{\rm NLTE}$} & 0.97 &  \ldots & 1 &  & 0.80 &  \ldots & 1 &  &   \ldots &   \ldots & 0 & &  \ldots &  \ldots & 0 \\
C\,{\sc i}       	       & $-$0.02 &  \ldots  & synth &  & 0.09 &   \ldots  & synth &  & $-0.27$  &   \ldots  & synth& & $-0.16$ &  \ldots  & synth \\ 
Na\,{\sc i}\rlap{$^{\rm LTE}$}& 0.00 & 0.02 & 4 &  & 0.08 & 0.23 & 4 &  & $-$0.02 & 0.06 & 4 & & 0.54 & 0.09 & 4\\ 
Na\,{\sc i}\rlap{$^{\rm NLTE}$}& $-$0.31 & 0.18 & 4 &  & $-0.17$ & 0.14  & 4 &  & $-0.28$ & 0.21 & 4 & & 0.10 & 0.22 & 4\\ 
Mg\,{\sc i}       	       & 0.43 & 0.06 & 3 & & 0.50 & 0.13 & 4 & & 0.50 & 0.04 & 4 & & 0.53 & 0.21 & 4  \\ 
Si\,{\sc i}       	       & 0.52  & 0.20 & 2 &  &  \ldots &  \ldots & 0 &  & 0.59 &  \ldots & 1 & &0.50  & 0.19 & 2 \\ 
Ca\,{\sc i}       	       & 0.40 &  0.33 & 18 &  & 0.38 &  0.23 & 12 &  & 0.31  & 0.24 & 14 & & 0.43  &0.23  & 15 \\ 
Sc\,{\sc ii}      	       & $-$0.02 &  0.12 &  7 &  & $-0.12$ & 0.14 & 7 &  & 0.00 & 0.07 & 7 & & $-0.04$ & 0.12 & 7\\ 
Ti\,{\sc i}       	       & 0.19 & 0.23 & 14 &  & 0.32  & 0.13  & 11  &  & 0.35 & 0.16 & 14 & & 0.32  & 0.15 & 12\\ 
Ti\,{\sc ii}      	       & 0.32 & 0.17 & 12 &  & 0.30  & 0.10  & 10 &  & 0.33 & 0.15 & 12 & & 0.19  & 0.09 & 11\\ 
V\,{\sc i}        	       & $-0.22$ &  \ldots  & 1 &  &  $-0.03$ &  \ldots & 1 & &  $-0.16$ &  \ldots & 1& & $-0.31$ &  \ldots & 1 \\ 
V\,{\sc ii}        	       &  \ldots &  \ldots  & 0 &  &  \ldots &  \ldots & 0 & & 0.10 &  \ldots & 1&  & $-0.12$ &  \ldots  & 1 \\ 
Cr\,{\sc i}       	       & $-$0.28 & 0.22 & 6 &  & $-0.38$ & 0.28 & 6 &  & $-0.21$ & 0.10 & 5 & & $-0.16$ & 0.09 & 6\\ 
Cr\,{\sc ii}       	       & $-$0.06 &  \ldots & 1 &  & 0.03 &  \ldots & 1 &  & $-0.02$ &  \ldots  & 1 & & $-0.11$ &  \ldots & 1 \\ 
Mn\,{\sc i}       	       & $-0.66$ & 0.25 & 6  &  & $-0.46$ & 0.38 & 6 &  & $-0.42$ & 0.26 & 6 & &$-0.56$ & 0.24 & 6 \\ 
Co\,{\sc i}                   & 0.03 & 0.10 & 4 &  & 0.00 & 0.29 & 4 &  & 0.21 & 0.24 & 4 & & 0.03 & 0.17 & 4\\ 
Ni\,{\sc i}       	       & 0.02 & 0.12 & 5 &  &  $-0.03$ & 0.18  & 2 &  & 0.16 & 0.09 & 3 & & $-0.09$ & 0.15 & 3 \\ 
Zn\,{\sc i}       	       & $-0.03$ & 0.00 & 2 &  &  0.37 & 0.08  & 2 &  & 0.16 & 0.16 & 2 & & 0.12 & 0.02 & 2\\ 
Sr\,{\sc ii}                  & $-0.08$ & 0.08 & 2 &  &  0.20 & 0.05 & 2 &  & $-0.10$ & 0.12 & 2 & & 0.48 & 0.04 & 2 \\ 
Y\,{\sc ii} 	  	       & $-0.08$ & 0.13 & 8 &  & $-0.10$ & 0.20 & 8 &  & $-0.20$ & 0.19  & 3 & & 0.11 & 0.12 & 6 \\ 
Zr\,{\sc ii}       	       & 0.14 &  \ldots  & synth &  & 0.31 &  \ldots  & synth &  & 0.17 &  \ldots & synth & & 0.32 &  \ldots & synth \\ 
Ba\,{\sc ii}                 & $-0.11$ & 0.19 & 5 &  & $-0.27$ & 0.17 & 5 &  & $-0.26$ & 0.12 & 5 & & 0.35 & 0.09 & 6 \\ 
La\,{\sc ii}	  	       & $-0.37$ & 0.16  & 3  &  & $-0.23$ & 0.15 & 3 &  & $-0.39$  &  \ldots & 1 & & 0.05 & 0.13 & 3 \\ 
Nd\,{\sc ii}	  	       & 0.09 & 0.13 & 2 &  & 0.37 &  0.25 & 3 &  & $-0.01$ &  \ldots  & 1 & & 0.35 & 0.07  & 2 \\ 
Eu\,{\sc ii}	  	       & 0.00 &  \ldots & 1 &  & $-0.01$ &  \ldots & 1 &  & 0.05 &  \ldots & 1 & & 0.02&   \ldots  & 1 \\
 \hline
 \hline
 \end{tabular}
\tablefoot{
\tablefoottext{a}{Ionised  species are given relative to Fe\,{\sc ii}. Abundance ratios are listed relative to iron, except for Fe\,{\sc i}  and Fe\,{\sc ii} (relative to H) and  Li. }}
\end{table*}
\begin{table*}[!htb]
\caption{Abundance results for the CH star, the CEMP-s star, and the metal-rich bulge giant.}             
\centering          
\begin{tabular}{crcrcrcrcrcr}     
\hline\hline       
 {} & {}  &  {10464}  & {}  & {}  &  {} &  {27793} & {}   & {}  &  {} &  {14135} & {}  \\
\cline{2-4}\cline{6-8}\cline{10-12}
\raisebox{1.5ex}[-1.5ex]{Element\tablefootmark{a}} &  {[X/Fe]}  &  {$\sigma$} &  {N}  &  {} &   {[X/Fe]} &  {$\sigma$} &  {N}  
 &  {} &   {[X/Fe]} &  {$\sigma$} &  {N} \\
 \hline
Fe\,{\sc i}        	       & $-1.53$ & 0.20 & 94 &  & $-2.52$ & 0.15 & 54  & & $-$0.01 & 0.22 & 145 \\ 
Fe\,{\sc ii}       	       & $-1.52$ & 0.10 & 8 &  & $-2.52$ & 0.11  & 4  & & \phantom{$-$}0.00 & 0.38 & 19 \\ 
C\,{\sc i}       	       &   0.41 &  \ldots  & synth &  & 1.44 &  \ldots  & synth   & & 0.00 &  \ldots  & synth \\ 
Na\,{\sc i}\rlap{$^{\rm LTE}$} &   0.54 & 0.06  & 4  &  & 0.51  & 0.17 & 4  & & 0.20 & 0.23  & 3 \\ 
Na\,{\sc i}\rlap{$^{\rm NLTE}$}&   0.01  & 0.12  & 4 &  & 0.14  & 0.22  & 4  & & 0.03 & 0.24 & 3 \\ 
Mg\,{\sc i}       	       &   0.49 & 0.15 & 4 & & 0.63 & 0.40 & 4 & & 0.17 & 0.08 & 6 \\ 
Si\,{\sc i}       	       &   0.58 & 0.08  & 2 &  &  \ldots  &  \ldots & 0  && 0.21 & 0.26 & 19 \\ 
Ca\,{\sc i}       	       &   0.19 & 0.22  & 17 &  & 0.42 & 0.28 & 12 & & $-0.01$ & 0.26 & 11\\ 
Sc\,{\sc ii}      	       &   0.00 & 0.12 & 4 &  & 0.13 & 0.11  & 2 & & 0.05  & 0.24 & 5 \\ 
Ti\,{\sc i}       	       &   0.37 & 0.27  & 14 &  & 0.42 & 0.14 & 6 & & 0.03 & 0.25 & 31 \\ 
Ti\,{\sc ii}      	       &   0.09 &  0.10 & 12 &  & 0.39 & 0.06 & 8 & & $-0.05$ & 0.36 & 12 \\ 
V\,{\sc i}        	       & $-0.18$ &  \ldots & 1 & & \ldots & \ldots & 0&  & 0.05 & 0.15 & 6 \\ 
Cr\,{\sc i}       	       & $-0.28$ & 0.10 & 6 &  & $-0.22$ & 0.27 & 6 & & $-0.04$ & 0.28  & 11 \\ 
Cr\,{\sc ii}       	       & $-0.11$ &  \ldots & 1 &  &  \ldots &  \ldots & 0 &  & 0.07 & 0.02 & 2 \\ 
Mn\,{\sc i}       	       & $-0.26$ & 0.29 & 6 &  & $-0.33$  & 0.19  & 4 & & $-0.01$ & 0.23 & 6\\ 
Co\,{\sc i}                  & 0.04 & 0.13 & 3 &  & 0.02 & 0.17 & 3 && $-0.13$ & 0.10 & 3 \\ 
Ni\,{\sc i}       	       & 0.03 & 0.20 & 5 &  & 0.08 & 0.17 &  3 && 0.08 & 0.25 & 36 \\ 
Zn\,{\sc i}       	       & 0.31 &  0.03 & 2 &  & 0.35 &  \ldots & 1 & &  \ldots &  \ldots & 0 \\ 
Rb\,{\sc i}                   & 1.29 &  \ldots & synth &  &  \ldots &  \ldots & 0 && 0.10 &  \ldots & synth \\ 
Sr\,{\sc i}                   & 0.59 &  \ldots & 1 &  & 0.59 &  \ldots & 1& & $-0.09$ &  \ldots & 1\\ 
Sr\,{\sc ii}                  & 0.84 & 0.05  & 2 &  & 0.58  & 0.16 & 2 &&  \ldots &  \ldots & 0 \\ 
Y\,{\sc ii} 	  	       & 0.46 & 0.14 & 3 &  & 0.51 & 0.27  & 5 & & 0.13 & 0.11 & 2 \\ 
Zr\,{\sc ii}       	       & 0.30 &  \ldots  & synth &  & 0.81 &  \ldots  & synth & &  \ldots &  \ldots & 0 \\ 
Ba\,{\sc ii}                 & 1.35 & 0.01 & 2  &  & 1.31 & 0.17  & 6 && 0.03 & 0.07 & 3 \\ 
La\,{\sc ii}	  	       & 0.92 & 0.20 & 3 &  & 1.29 & 0.13 & 3  &&  \ldots &  \ldots & 0  \\ 
Nd\,{\sc ii}	  	       & 1.13 & 0.08 & 2 &  & 1.18 & 0.02 & 2 & &  \ldots &  \ldots & 0 \\ 
Eu\,{\sc ii}	  	       & 0.37 &  \ldots & 1 &  &  \ldots &  \ldots & 0& &  \ldots &  \ldots & 0 \\
 \hline
 \hline
 \end{tabular}
\tablefoot{
\tablefoottext{a}{Ionised  species are given relative to Fe\,{\sc ii}. Abundance ratios are listed relative to iron, except for Fe\,{\sc i}  and Fe\,{\sc ii} (relative to H) and  Li. }}
\end{table*}
\subsection{Error analysis}
As a measure for the statistical errors on the derived abundances, we list in Tables~5 and 6 the 1$\sigma$ line-to-line scatter and number of lines, N, used
in the analysis, which leads to small random uncertainties in those elements with many suitable transitions such as Fe, Ca, and Ti.   
For more ill-defined cases, where only one line was measurable, we assessed a 1$\sigma$ error using the formalism of Cayrel (1988), which 
accounts for the local continuum noise and the width of the lines. 
In some cases, the resulting error for individual stars is large, up to 0.15 dex on, e.g., the Li, Si, or Nd abundances when derived from the 6707, 5948, 4061\AA~transitions
alone. Typically, this random error is on the order of 0.08 dex on average. Eu abundances could only be derived from the 4129\AA-line to an accuracy of 0.08 dex.

In order to quantify the systematic errors on the final   abundance ratios we used the standard approach of computing nine new atmosphere models 
with altered stellar parameters. Thus, each of  (T$_{\rm eff}$, log\,$g$, $\xi$, [$M$/H], [$\alpha$/Fe]) was varied by its typical uncertainty, estimated above 
as ($\pm$50 K, $\pm$0.15 dex, $\pm$0.10 km\,s$^{-1}$, $\pm$0.1 dex, $\pm$0.1 dex). 
To test  the influence of the atmospheric $\alpha$-enhancement, we re-ran our analysis using the solar-scaled opacity distributions, ODFNEW, which mimics  an uncertainty 
in the model [$\alpha$/Fe] ratio of 0.4 dex. 
New abundance ratios were then determined and we list in 
Table~7  the deviation in $\log\,\varepsilon$ from those given in Tables~5,6, derived using  the unaltered stellar parameters. 
This exercise was carried out  for the coolest and the hottest stars in our metal-poor sample, viz. 17221 and 10464.
\begin{table*}[htb!]
\caption{Systematic error analysis for the red giant 17221 and the HB star 10464.}
\centering          
\begin{tabular}{rccccrc|ccccrc}     
\hline\hline       
 {} &  {$\Delta$T$_{\rm eff}$} &  {$\Delta\,\log\,g$} &  {$\Delta\xi$} &  {$\Delta$[M/H]} & {}  &  {}  & 
                        {$\Delta$T$_{\rm eff}$} &  {$\Delta\,\log\,g$} &  {$\Delta\xi$} &  {$\Delta$[M/H]} & {}  &  {}   \\
\raisebox{1.5ex}[-1.5ex]{Ion}  &  {$\pm$50\,K}  &  {$\pm$0.15\,dex}  &  {$\pm$0.1\,km\,s$^{-1}$}  & {$\pm$0.1\,dex} & \raisebox{1.5ex}[-1.5ex]{ODF} & \raisebox{1.5ex}[-1.5ex]{$\sigma_{\rm tot}$} &  
                                                          {$\pm$50\,K}  &  {$\pm$0.15\,dex}  & {$\pm$0.1\,km\,s$^{-1}$}  & {$\pm$0.1\,dex} &  \raisebox{1.5ex}[-1.5ex]{ODF} & \raisebox{1.5ex}[-1.5ex]{$\sigma_{\rm tot}$}  \\
\cline{2-7}\cline{8-13}
& \multicolumn{6}{c}{17221} &  \multicolumn{6}{c}{10464} \\
 \hline
Fe\,{\sc i}  & $\pm$0.06 & $\mp$0.02 & $\mp0.03$ & $<$0.01 & 0.01 & 0.07 &  $\pm$0.05  & $<$0.01 & $\mp$0.03  & $<$0.01 & 0.01 & 0.06  \\ 
Fe\,{\sc ii} & $<$0.01 & $\pm$0.05 & $\mp$0.03 & $<$0.01 & $-0.01$ & 0.06  & $<$0.01 & $\pm$0.06 & $\mp$0.03 & $<$0.01 & $-0.01$  & 0.07  \\ 
A(Li)  & $\pm$0.06 & $<$0.01 &  $<$0.01 &  $<$0.01 &  $<$0.01  & 0.06  &  \ldots &  \ldots  &   \ldots &   \ldots &  \ldots  &    \ldots  \\
C\,{\sc i}   & $\pm$0.11 & $\mp$0.08 & $<$0.01 & $\mp$0.09 & $-0.18$ & 0.17& $\pm$0.10 & $\mp$0.06 & $<$0.01 & $\mp$0.08  & $-0.04$ & 0.14  \\
Na\,{\sc i}  & $\pm$0.06 & $\pm$0.02 & $\pm$0.02 & $<$0.01 &  $<$0.01 & 0.06 &  $\pm$0.05 & $\mp$0.01  & $\mp$0.03& $\mp$0.01& $<$0.01 & 0.06   \\ 
Mg\,{\sc i}  &  $\pm$0.05 & $\mp$0.03 & $\mp$0.02 &   $<$0.01 & $<$0.01   & 0.06 &    $\pm$0.04  &	$<$0.01  &    $\mp$0.03 &    $<$0.01 &    $<$0.01 &   0.05   \\ 
Si\,{\sc i}  &  $\pm$0.04 & $\mp$0.02 & $\mp$0.01 &   $<$0.01 & $<$0.01   & 0.05 &    $\pm$0.02  &	$<$0.01  &    $\mp$0.01 &    $<$0.01 &    $<$0.01 &   0.03  \\ 
Ca\,{\sc i}  &  $\pm$0.04 & $\mp$0.02 & $\mp$0.01 & $\pm$0.01 & $<$0.01   & 0.05 &    $\pm$0.03  &	$<$0.01  &    $\mp$0.02 &    $<$0.01 &    $-$0.01 &   0.04  \\ 
Sc\,{\sc ii} & $\pm$0.02 & $\pm$0.05 & $\mp0.02$ & $<$0.01 & $-$0.01 & 0.06 & $\pm$0.03 & $\pm$0.06 & $\mp$0.04 & $\pm$0.01& $-0.01$ & 0.08 \\ 
Ti\,{\sc i}  &  $\pm$0.07 & $\mp$0.02 & $\mp$0.01 &   $<$0.01 &    0.01   & 0.07 &    $\pm$0.06  &	$<$0.01  &    $\mp$0.02 &    $<$0.01 &    $-$0.01 &   0.06  \\ 
Ti\,{\sc ii} &  $\pm$0.02 & $\pm$0.04 & $\mp$0.01 &   $<$0.01 &    0.01   & 0.05 &    $\pm$0.02  &    $\pm$0.05  &    $\mp$0.02 &    $<$0.01 &    $-$0.02 &   0.06  \\ 
V\,{\sc i}   &  $\pm$0.07 & $\mp$0.02 & $<$0.01 &   $<$0.01 &    0.02   & 0.07 &    $\pm$0.06  &    $\pm$0.01  &    $<$0.01 &    $<$0.01 &    $<$0.01 &   0.06  \\ 
V\,{\sc ii}  &  $\pm$0.01 & $\pm$0.04 & $\mp$0.01 &   $<$0.01 &    0.01   & 0.05 &    $\pm$0.02  &    $\pm$0.05  &    $\mp$0.03 &    $<$0.01 &    $-$0.01 &   0.06  \\ 
Cr\,{\sc i}  &  $\pm$0.06 & $\mp$0.01 & $\mp$0.01 &   $<$0.01 &    0.01   & 0.06 &    $\pm$0.05  &    $\mp$0.01  &    $\mp$0.02 &    $<$0.01 &    $<$0.01 &   0.06 \\ 
Cr\,{\sc ii} &  $\mp$0.01 & $\pm$0.04 & $<$0.01 &   $<$0.01 &    0.01   & 0.04 &    $<$0.01  &    $\pm$0.05  &    $\mp$0.02 &    $<$0.01 &    $-$0.01 &   0.06  \\ 
Mn\,{\sc i}  & $\pm$ 0.06& $\mp$0.03 & $\mp$0.03 & $<0.01$ &  0.02 & 0.08 & $\pm$0.06 & $\mp$0.01 & $\mp0.04$ & $<0.01$ & 0.01 & 0.07 \\ 
Co\,{\sc i}  &$\pm$0.07 & $\mp$0.03 & $\mp$0.05  & $<$0.01 & 0.02 & 0.09 & $\pm$0.06 & $<$0.01 & $\mp$0.04 & $<$0.01 & $<$0.01 &  0.07  \\ 
Ni\,{\sc i}  &  $\pm$0.06 & $\mp$0.02 & $\mp$0.02 &   $<$0.01 &    0.02   & 0.07 &    $\pm$0.05  &    $\pm$0.01  &    $\mp$0.02 &    $<$0.01 &    $<$0.01 &   0.05  \\ 
Zn\,{\sc i}  &  $\pm$0.02 & $\pm$0.02 & $<$0.01 &   $<$0.01 & $<$0.01   & 0.03 &    $\pm$0.03  &    $\pm$0.01  &    $\mp$0.02 &    $<$0.01 &    $-$0.01 &   0.04  \\
Rb\,{\sc i} &   \ldots &   \ldots &   \ldots &    \ldots &   \ldots &   \ldots & $\pm$0.03  & $<$0.01 & $<$0.01  & $<0.01$  & $<$0.01  & 0.04   \\
Sr\,{\sc i} &   \ldots &   \ldots &   \ldots &    \ldots &   \ldots &   \ldots & $\pm$0.05 & $\mp$0.01 & $\mp$0.01 & $<$0.01 &$<$0.01  & 0.05  \\
Sr\,{\sc ii} & $\pm$0.04 & $<0.01$ & $\mp$0.06 & $<$0.01 & $<0.01$ & 0.07 & $\pm0.04$ & $\pm0.03$ & $\mp$0.02 & $\pm$0.01 & $-0.02$ & 0.06 \\
Y\,{\sc ii}  &  $\pm$0.03 & $\pm$0.04 & $\mp$0.02 &   $<$0.01 &    0.01   & 0.05 &    $\pm$0.02  &    $\pm$0.05  &    $\mp$0.04 &    $<$0.01 &    $-$0.01 &   0.07  \\ 
Zr\,{\sc ii} &  $\pm$0.03 & $\pm$0.04 & $\mp$0.01  & $<$0.01    & $<$0.01      &   0.05 &    $\pm$0.03   & $\pm$0.05      & $\mp$0.03    & $\pm$0.01    & $-0.02$   & 0.07  \\
Ba\,{\sc ii} & $\pm$0.03 & $\pm$0.04 & $\mp$0.04 & $<$0.01 & $-0.01$ & 0.07 & $\pm$0.05 & $\pm$0.04 & $\mp$0.06 & $<$0.01 & $-0.01$ & 0.08	\\ 
La\,{\sc ii} &  $\pm$0.03 & $\pm$0.04 & $<$0.01 &   $<$0.01 &    0.01   & 0.05 &   $\pm$0.03  &    $\pm$0.05  &    $\mp$0.07 &    $<$0.01 &	 $-$0.01 &   0.09  \\ 
Nd\,{\sc ii} &  $\pm$0.03 & $\pm$0.04 & $<$0.01 &   $<$0.01 &    0.01   & 0.05 & $\pm$0.04  &    $\pm$0.05  &    $\mp$0.05 &    $<$0.01 &    $-$0.01 &   0.08 \\ 
Eu\,{\sc ii} &  $\pm$0.03 & $\pm$0.04 & $<$0.01 &   $<$0.01 &    $-$0.01   & 0.05 & $\pm$0.03 & $\pm$0.05 & $<$0.01 & $<$0.01 & $-0.01$ & 0.06 \\
 \hline
 \hline
 \end{tabular}        	 
\end{table*}

Here, the largest systematic error source is T$_{\rm eff}$, while 
metallicity and the opacity distributions generally  have a neglibible influence, with the exception of carbon, where the latter parameters dominate.
As an upper, conservative limit ignoring possible covariances between the stellar parameters (e.g., McWilliam 1995), we summed up all contributions in quadrature. 
Since a typical error on the observed [$\alpha$/Fe] ratio is 0.1 dex, we only transfer 1/4 of the ``ODF'' uncertainty into the final error budget. 
The final, total systematic uncertainty values are listed in  the columns $\sigma_{\rm tot}$ in Table~7.
\subsection{Iron abundance}
With the exception of the $s$-process enhanced HB star 10464 (at $-1.5$ dex) and the regular, metal-rich bulge star at Solar metallicity, 
all of our targets are ``very metal-poor stars'' (Beers \& Christlieb 2005), i.e.,  below [Fe/H]$< -2$ dex, 
 reaching as low as $-2.66$ dex. 
 
Metal-poor stars ``in'' the bulge have been reported before, albeit in very low numbers: 
Kunder et al. (2012) confirmed the well-known vertical metallicity gradient in the bulge in that 
lower-latitude  ($-8\degr$) fields are more metal rich, and only a hint of a population of metal-poor stars could be detected.
Garc{\'{\i}}a P{\'e}rez et al. (2013) found two stars marginally below $-2$ dex, which constitutes 0.1\% of their large sample from the APOGEE. 
Howes et al. (2014) reported on four giants at $-2.7$ to $-2.5$ dex near $|b|\sim8\degr$--$9\degr$, and Schlaufman \& Casey (2015) devised 
a promising photometric selection method, from which {  Casey \& Schlaufman (2015) spectroscopically confirmed three stars towards the bulge that lie between $-3$ and $-2.7$ dex}.
One of the most comprehensive medium-resolution spectroscopic MDFs of the bulge was gathered by Ness et al. (2013a).
Their data indicate a complex mix of up to five populations throughout the bulge, with a prominent peak at +0.3 dex. 
However, only  0.11\% of their 14,000-star sample within 3.5 kpc  falls below $-$2 dex, reaching as low as $-2.8$ dex. 
This metal-poor tail gains importance towards more negative latitudes ($b \sim -10\degr$), i.e., further from the plane, 
which already  hints at an increasing overlap with underlying halo. 
In fact, Ness et al. (2013a) identify this low-number, metal-poor component with the inner Galactic halo, as later confirmed by Ness et al. (2013b) 
as a slowly rotating spheroidal population. 
All these lines of evidence illustrate the difficulty in telling metal-poor stars as members of the genuine, old bulge from an underlying
inner halo population and, based on the low metallicities of our targets alone, no firm association with either component could be achieved. 
\subsection{Light elements: C, Li, Na}
Li could only be measured from the resonance line at 6707\AA~ in two of the red giants, with EWs of 15 and 18 m\AA. 
The abundances indicate the strongly depleted levels below A(Li)$<1$,  as is expected for these evolved stars, since their interiors favour the 
downward transport and rapid destruction of this fragile element (e.g., Bonsack 1959; Iben 1965; Lind et al. 2009).

We determined carbon abundances in a $\chi^2$-sense from the 
$A^2\Delta - X^2\Pi$  G-band near 4320~\AA.
Our synthesis used a line list kindly provided by B. Plez (priv. comm.). 
The typical uncertainty of the fit lies at $\sim$0.15 dex  and is  driven by uncertainties in the continuum placement.
As Fig.~1 (top panel) indicates, four of the stars show [C/Fe] ratios that are fully consistent with metal-poor halo stars, while 
two stars, 10464 and 27793, show elevated C abundances. 
At [Fe/H]=$-2.52$ and [C/Fe]=1.44 dex, the latter provides a typical example of a carbon-enhanced metal-poor (CEMP) star (Beers \& Christlieb 2005). 
The origin of its carbon over-abundance is most likely the mass transfer from the envelope of 
an asymptotic giant branch (AGB) star binary companion (now a white dwarf; e.g., Aoki et al. 2007; Sneden et al. 2008;
Hansen et al. 2014), consistent with the observed very strong enrichment of $s$-process elements (Sect.~4.6). 
This star is discussed in further detail in Sect.~5.2. %
\begin{figure}[!htb]
\begin{center}
\includegraphics[angle=0,width=1\hsize]{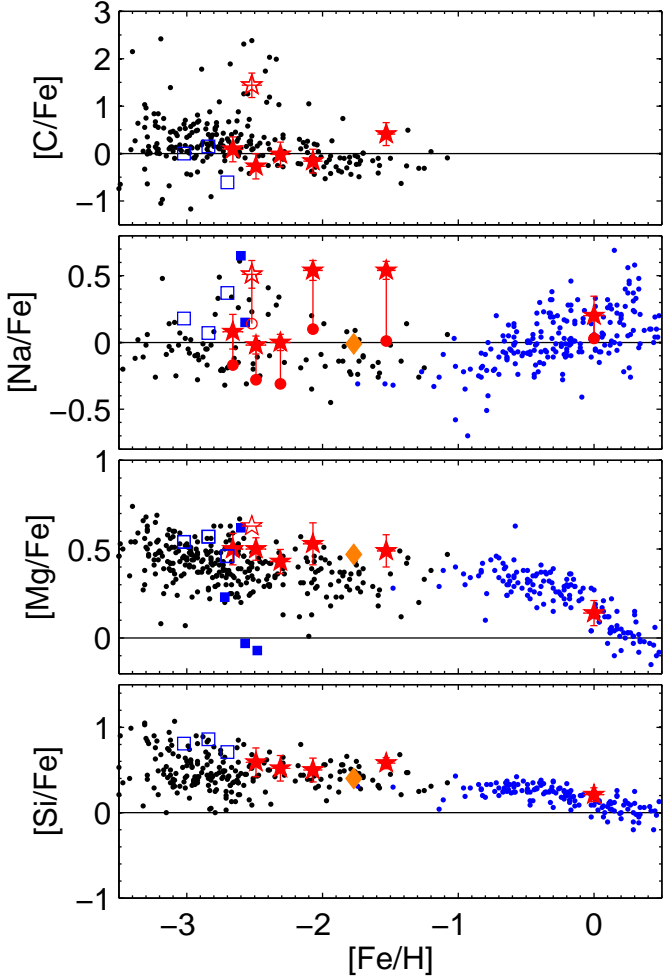}
\end{center}
\caption{Abundance results for C, Na, Mg, and Si. Our results are shown as red star symbols, where the open star singles out the CEMP-s star 27793. 
For Na, NLTE abundances are also shown by red circles. Error bars account for random and systematic uncertainties. 
We also include the metal-poor bulge candidates of Howes et al. (2014; filled blue squares) { and Casey \& Schlaufman (2015; open blue squares)}, 
the $r$-process enriched bulge star from Johnson et al. (2013b; orange diamond), and  blue points designate the bulge sample of Johnson et al. (2012, 2014).
Finally, black dots are halo stars from Roederer et al. (2014), while black triangles
are upper limits from that study. Note that their sample extended as metal-poor as [Fe/H]=$-4.6$ dex, 
but we truncated the axis for better readability.}
\end{figure}

Since the commonly used Na doublets near 5688 and 6154\AA~were undetectable in our metal-poor sample, we resorted 
to  the prominent Na D lines, provided they were not too strong and saturated, and the near-IR doublet at 8183, 8194\AA.  
In either case we assured that the stellar lines were well separated from ISM absorbers and telluric contamination. 
The derived abundances were corrected for departures from LTE using predictions from Lind et al. (2011)\footnote{Taken from the authors' web-based database, {\tt www.inspect-stars.com}.}
and in Tables~5,6 and Fig.~1 we show both LTE and NLTE values. 
While the Na abundances in our stars appear high, they are still compatible with the bulk of halo reference stars at similar metallicity. 
\subsection{$\alpha$-elements: Mg, Si, Ca, Ti}
\begin{figure}[!tb]
\begin{center}
\includegraphics[angle=0,width=1\hsize]{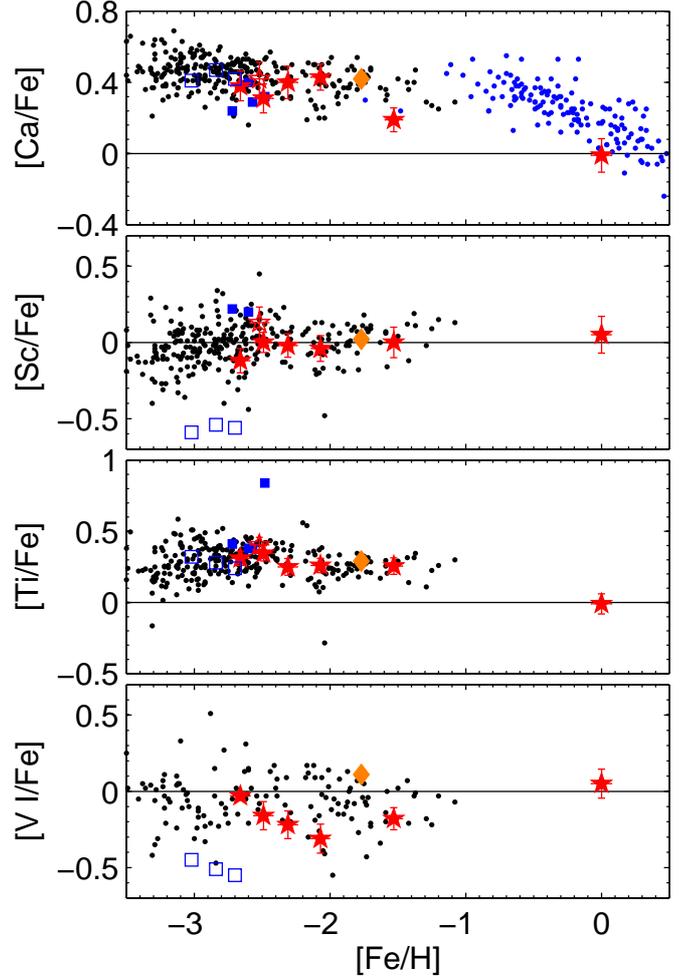}
\end{center}
\caption{Same as Fig.~1, but for Ca, Sc, Ti, and V.}
\end{figure}
For the $\alpha$-elements Mg, Ca, and Ti, a wealth of absorption lines are detectable in the metal-poor stars' spectra, while 
for Si we were only able to measure one or two of the lines of moderate strength at 5948 and 7423\AA. 
The literature data for Mg we used for comparison in Fig.~1 were not corrected for NLTE; 
Andrievsky et al. (2010) computed NLTE corrections of the order 0.2 -- 0.4 dex (in the sense LTE$-$NLTE)  for halo stars of similar stellar 
parameters, albeit at metallicities lower  by $\sim$1 dex.

{Fig.~1 indicates that our [Mg/Fe] is higher (by $\sim$0.15 dex) than the Roederer et al. (2014) halo sample, at a mean of 0.51 dex and a low 1$\sigma$ dispersion of 0.07 dex. It is also important to note that
$r$-process rich bulge star from Johnson et al. (2013) resembles our metal-poor bulge sample more than it does the halo stars. 
Differences in the $gf$-values between Roederer's and our sample cannot account for more than 0.02 dex in the abundances, while the use of damping constants of Barklem et al. (2000)\footnote{These are, however, only 
available for a subset of our lines.} vs. the standard Uns\"old (1955) approximation,  leads to Mg abundances lower by $\sim$0.06 dex.}

Ionisation balance is also fulfilled for Ti in that  the abundances of its neutral and ionised species
 agree to within, with 0.05 dex(1$\sigma$ dispersion of 0.11 dex).

Most $\alpha$-element abundances in the regular sample stars follow the distributions on the metal-poor halo plateau in lockstep remarkably well.
In particular we find mean values for [Si,Ca,Ti/Fe] of 0.55, 0.36, and 0.33 dex, respectively, if we exclude the metal-rich star 14135.
Also, the low dispersion amongst the sample is noteworthy, where we observe a 1$\sigma$-scatter in these elements of 
0.07, 0.04, 0.09, and 0.08 dex.  
This low scatter pertains even when including the two stars with large overabundances of C and  the $s$-process elements, which merely reflects that 
the production of the $\alpha$-elements is fully detached from the AGB-nucleosynthesis involved in the pollution of the CEMP-$s$ and CH stars's
atmospheres. 

This conformity was not seen amongst the four metal-poor bulge candidates analysed by  Howes et al. (2014), where one star has an abnormally 
high Ti abundance (see the blue squares in Figs.~1--5), { while the most metal-poor candidates by Casey \& Schlaufman (2015) show very low scatter, as for our stars}. 
Furthermore, { Howes' et al. (2014) spectra} revealed a broad range in Mg abundances, spanning across 0.7 dex from sub-solar values to several dex above the halo plateau.

One might argue that the metal-rich giant 14135 could be associated with the Galactic thin or thick disk, given its high, solar metallicity. 
As Fig.~1 indicates, however, its  [Mg/Fe] ratio is elevated, at 0.17 dex,  and is consistent with Johnson's et al. (2014) sample that shows [Mg/Fe]=0.14 dex around solar metallicity, 
providing chemical evidence for an association with the metal-rich bulge (see also Fulbright et al. 2007). 
\subsection{Scandium: no evidence for Population~III enrichment}
For Sc, our seven target stars display abundances that lie at  the Solar value (at a mean and 1$\sigma$ dispersion of 0.00 and 0.08 dex, respectively), in line with the Galactic halo.
A number of studies have claimed $\sim$0.2 dex [Sc/Fe] enhancements in metal-poor stars of the
MW halo and thick and thin disks (e.g., Zhao \& Magain 1990; Nissen et al. 2000;
Carretta et al. 2002; Cohen et al. 2004; Brewer \& Carney 2006; Reddy et al. 2006).  However, it 
has been pointed-out that errors in log~$gf$ values and/or the treatment of hyperfine structure (e.g., Gratton \& Sneden 1991; 
Prochaska \& McWilliam 2000) may explain such putative [Sc/Fe] enhancements.

Because most of our stars scatter about the normal halo [Sc/Fe] trend with [Fe/H], the slightly elevated 
[Sc/Fe] enhancement to 0.13 dex for star 27793  cannot be due to spurious log~$gf$ values.  The Sc~II lines
in our program stars are weak and unsaturated, on the linear portion of the curve of growth, so any 
hyperfine splitting effects are very small and not likely to produce markedly different results for an individual star.
Likewise, we ascertained that none of the lines used in the CEMP star was blended with, generally strong, C$_2$ and CH lines. 

It is important to emphasise that 
the [Sc/Fe] ratios of our stars  are similar to those in outer MW halo stars 
(e.g., Gratton \& Sneden 1991; McWilliam 1995; Roederer et al 2014).  
This is in stark contrast to the 
extremely low [Sc/Fe] ratios predicted for massive Population III SNe by, e.g., 
Nomoto et al. (2006) and Heger \& Woosley (2010), { and to the recent observation  of such depletions in 
bulge stars by Casey \& Schlaufman 2015}.  
Simulations by, e.g., Diemand et al. (2005) and Tumlinson (2010) predicted that 
the fraction of these first, massive stars should peak in the bulge and inner, central halo (R$_{\rm GC}\la$3 kpc) of
the Galaxy. Even though our metal-poor stars are located in these central regions (note that three of the stars lie within 3 kpc of the Galactic centre; see Sect. 6.2), we thus find no 
evidence that they had been enriched by massive Population III  nucleosynthesis. 
\subsection{Fe-peak: V, Cr, Mn, Co, Ni, Zn}
\begin{figure}[!htb]
\begin{center}
\includegraphics[angle=0,width=1\hsize]{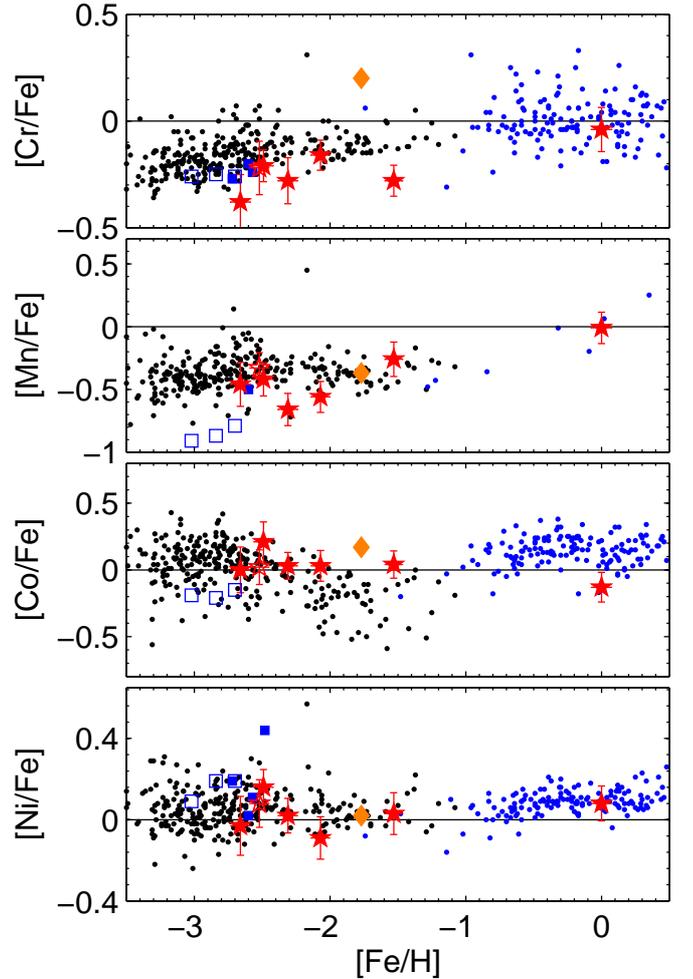}
\end{center}
\caption{Same as Fig.~1, but for Cr, Mn, Co, and Ni. Bulge data for Mn were taken from McWilliam et al. (2003).}
\end{figure}
Elements of the iron-peak are made by supernovae of both Type~Ia and Type~II (henceforth SNe~Ia and
SNe~II, respectively).  However, for the MW thin disk, at solar metallicity, it is thought that
roughly 2/3 of iron-peak elements are produced by SNe~Ia and 1/3 by SNe~II.  
The abundance ratios we derived in our sample of bulge stars bear little surprise 
in comparison with halo stars of similarly low metallicity and, as for the $\alpha$-elements,
we note a remarkable homogeneity in the iron-peak across the covered metallicity range.

Thus, we find sub-solar [X/Fe] abundance ratios for V, Cr, and Mn.
The sub-solar [Cr/Fe] values found here are consistent with the trend of low [Cr/Fe] in metal-poor
halo stars, initially found in the LTE abundance study of McWilliam et al. (1995), but also
seen in the results of Cayrel et al. (2004) and Roederer et al. (2014); notably, our [Cr/Fe]
values populate the lower edge of the trend of Roederer et al. (2014).  These LTE results follow
a seemingly consistent trend of declining [Cr/Fe] with decreasing [Fe/H], from solar metallicity,
as nicely shown by Cohen et al. (2004).

However, a NLTE analysis of Cr in metal-poor, warm dwarfs by Bergemann \& Cescutti (2010), led
them to conclude that the deficiencies found in LTE analyses could be explained by
NLTE effects, rather than genuine deficiencies in [Cr/Fe].  This conclusion is supported
by the lack of ionisation equilibrium obtained in this study: from our single Cr~II line we
find [Cr~I/Cr~II]=$-$0.19 dex here (similar to the typical NLTE correction
of Bergemann \& Cescutti 2010); thus, our [Cr~II/Fe~II] points lie close to the solar ratio.

Star 10464, while exhibiting a low [Cr/Fe] ratio, appears significantly below the general
trend of [Cr/Fe] versus [Fe/H], described above.  Star 10464 is a relatively hot, metal-poor
star, hotter than the rest of our sample, with T$_{\rm eff}$=5500 K; thus, it seems possible
that NLTE over-ionisation of Cr may be enhanced in this star, leading to it's lower than
expected LTE [Cr/Fe] ratio.  We further discuss star 10464 in Sect.~5.1. 

The fact that the [Mn/Fe] ratios in bulge stars follow the trend of the Galactic disks led McWilliam et al. (2003) to conclude  
that Mn is produced in SNe~Ia and SNe II with metallicity-dependent yields. Here we note a similarly good agreement between the [Mn/Fe] 
ratios in the Galactic halo and the metal-poor bulge candidates. 

While our [V~I/Fe~I] abundance ratios all lie below the solar value, at $-$0.18 dex, they fall
within the scatter of [V/Fe] ratios, from V~I lines, measured in the survey of
Roederer et al. (2014), which are on average sub-solar, near $-$0.15 dex.   However, 
abundances derived
by Roederer et al. (2014) from V~II lines are enhanced, relative to solar, by $\sim$0.2 dex;
the difference between V~I and V~II line abundances might be expected from NLTE over-ionisation of
neutral vanadium.  In this work we detected a single V~II line in two program stars,
yielding [V~II/Fe]=$-$0.02 dex; these two V~II detections could be explained by noise, or
a genuine enhancement of V~II over V~I due to NLTE over-ionisation of V~I.

On the other hand, the recently improved V~I and V~II line $gf$ values by Lawler et al. (2014) and 
Wood et al. (2014) that we employ in this work resulted in consistent [V/Fe] abundance ratios for the Sun and a metal-poor 
star: neutral and ionised line [V/Fe] ratios of $+$0.25 and $+$0.24 dex, respectively, were
found for HD84937 with [Fe/H]=$-$2.32.
Thus the new line $gf$ values  indicate a slight enhancement in [V/Fe] at low metallicity.  Clearly, the choice of $gf$
values is critical for determining the vanadium ionisation equilibrium and [V/Fe] at low
metallicity. Notwithstanding, our results appear consistent with previous studies.

Cobolt in all stars is solar to moderately elevated, although we do not trace the slightly
decreasing trend, towards higher metallicity, commencing at $\sim -2$ dex, seen in the
Roederer et al. (2014) data.  The [Co/Fe] trend in the Roederer et al. (2014) data roughly
follows the same trend found by McWilliam et al. (1995), and seen by Cayrel et al. (2004), 
but to higher [Fe/H], where the [Co/Fe] ratios continue lower.
Our Co enhancement is, however, driven by the CH star 
10464 with a [Co/Fe] value that is higher than seen in halo stars. Here, we carefully checked that none of the lines we used (at 3995, 4118, 4121\AA) suffered from   
blends with near-by CH- or C$_2$-bands.
NLTE corrections for Co are large:  Bergemann et al. (2010) estimate departures of Co (NLTE$-$LTE)  
of $+$0.4 dex on average  for dwarfs and up to $+$0.64 dex in [Co/Fe]  for the one metal-poor giant in their sample with parameters similar to our stars.
The maximum line-by-line deviation can even be higher, with corrections up to 0.9 dex for the case of the 4121\AA~line that we also employed in our analysis.  However, we note that applying NLTE corrections increases the [Co/Fe] ratios to well above the solar value,
in the same sense as the enhancements found By McWilliam et al. (1995) and Cayrel et al. (2004).

As is found throughout all main Galactic components Ni is scattered around the value solar (at a mean [Ni/Fe] of 0.03 dex with a 1$\sigma$ scatter of 0.09 dex), which merely reflects the common production 
of Ni and Fe in the SNe Ia. 
Finally, Zn abundances were derived from two lines of moderate strength at 4722 and 4810\AA. While 
Zn is often associated with the heaviest Fe-peak elements to be formed in core-collapse SNe, an increase of the [Zn/Fe] ratio with decreasing metallicity (e.g., Johnson \& Bolte 2001; Cayrel et al. 2004; Honda et al. 2011) 
has prompted suggestions that Zn could also have contributions from neutron-capture processes in the main $s$-process (e.g., Baraffe et al. 1992)
or in energetic hypernovae (e.g., Umeda \& Nomoto 2002). This trend is also visible in our data below an [Fe/H] of $-2.3$ dex. 
\subsection{Neutron-capture elements: Sr, Y, Zr, Ba, La, Eu}
All of the elements formed in the slow ($s$-) and rapid ($r$-) neutron-capture processes are fully compatible with the halo distributions (Figs.~4,5), 
except for the two stars that are strongly enhanced in $s$-process elements (10464 and 27793). 
\begin{figure}[!tb]
\begin{center}
\includegraphics[angle=0,width=1\hsize]{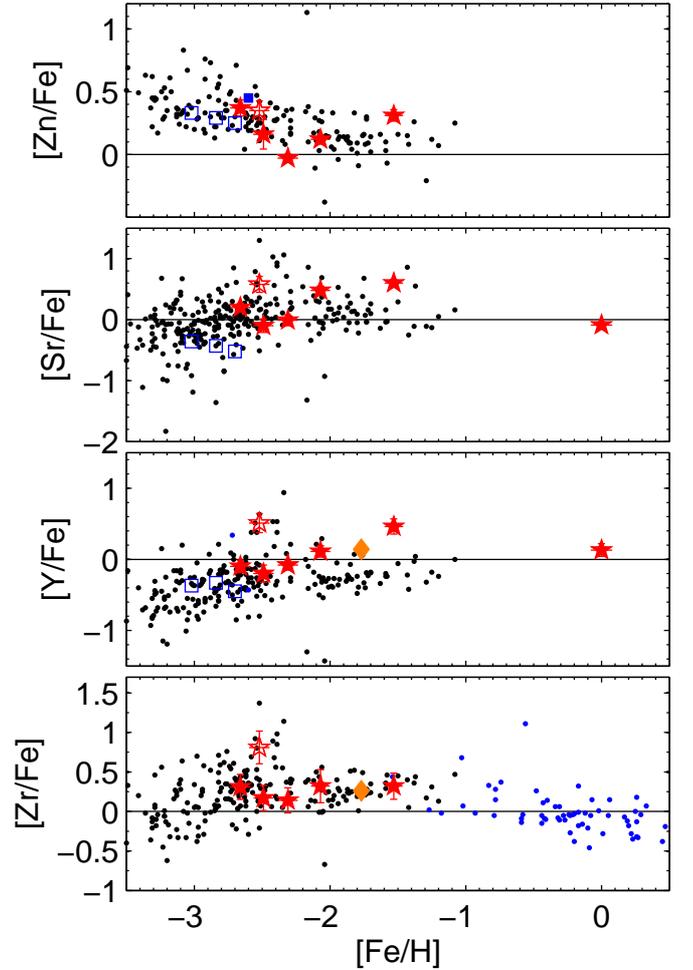}
\end{center}
\caption{Same as Fig.~1, but for Zn, Sr, Y, and Zr. Here, the comparison sample for the bulge has been taken from Johnson et al. (2012).}
\end{figure}
\begin{figure}[!htb]
\begin{center}
\includegraphics[angle=0,width=1\hsize]{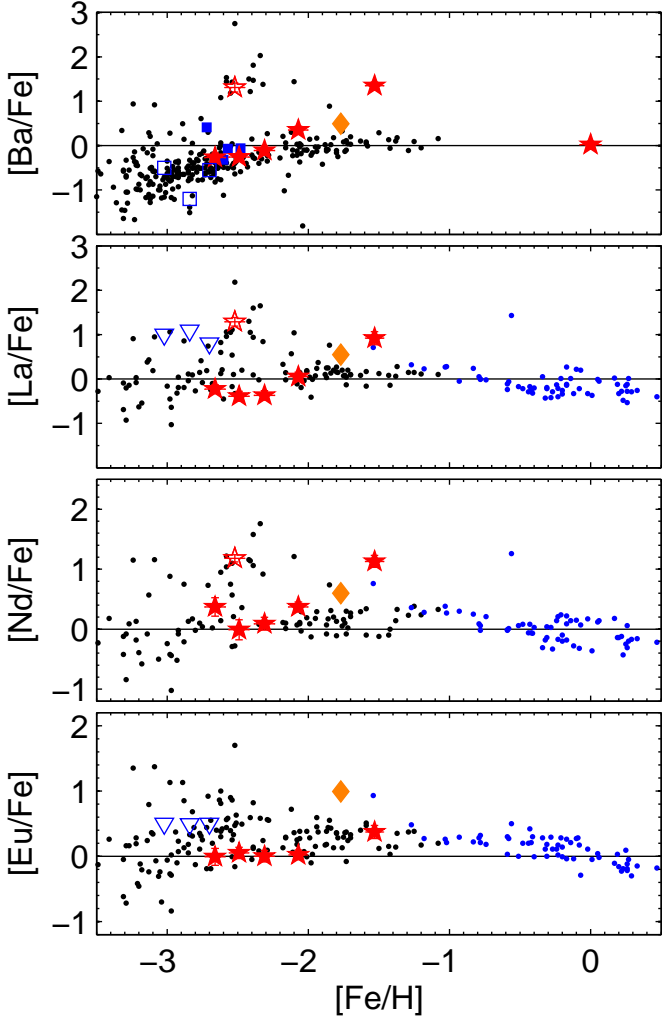}
\end{center}
\caption{Same as Fig.~4, but for Ba, La, Nd, and Eu. { Here, open blue triangles indicate that upper limits on the measurements from Casey \& Schlaufman 2015.}}
\end{figure}

For the case of Zr we obtained abundances from spectral synthesis of  three well-defined, unblended lines (4161, 4208, 4496\AA). 
Their best-fit abundances agreed well to within 0.15 dex, which we adopt as statistical error on the Zr/Fe ratios. 
We determined Sr abundances from the two resonance lines at  4077 and 4215\AA, which  are generally strong and yield consistent results. 
NLTE corrections to Sr were calculated by  Hansen et al. (2013) and are on the order of 0.07 dex, albeit with a sign that is dependent on the 
exact stellar parameters. 

The Ba abundances we found for our sample strikingly follows the trend of metal-poor halo stars, except for the 
outlying CEMP-$s$ star 27793 and the $s$-rich star 10464, which both show strong enhancements, likely due to 
mass transfer of the $s$-process enhanced material from an AGB companion (Sect.~5). 
 Andrievsky et al. (2009) computed NLTE corrections for Ba for the same sample as for Sr and estimate corrections  0.25 dex with  opposite signs for the HB stars. 
Note, however, that these levels of corrections are probably inappropriate for the strongly enhanced stars. 

Values for the [Eu/Fe] abundance ratio could only be derived from the 4129\AA-line, since all other possible transitions are very weak and/or heavily blended (Mashonkina \& Christlieb 2014; Hansen 
et al. 2015). As a result, we find [Eu/Fe] ratios that are Solar throughout the sample, with only a mild indication of an enhancement in the CH star 10464. 
This mean value of [Eu/Fe] for our stars, at 0.1 dex (1$\sigma$ scatter of 0.16 dex), is surprisingly low, contrasting with the typical halo
enhancement near $+$0.4 dex seen in the Roederer et al (2014) data, and found by Fulbright
(2000; see also Woolf, Tomkin \& Lambert 1995). 
The low [Eu/Fe] ratios may be due to a genuine deficiency of r-process elements, perhaps due to
fewer sources. Alternatively, most of the stellar atmosphere material in our program stars may
have experienced slow neutron-capture; note that, due to its large neutron-capture cross section,
a very mild s-processing could drastically reduce the Eu abundance without greatly affecting the
abundances of other s-process elements.  Whatever the explanation, we might expect similar deficiencies
of other r-process elements in these stars.

The [Ba/Eu] ratio of the MW halo, in Fig.~6, shows a clear trend of increasing
$s$-process fraction with increasing [Fe/H] (see also Simmerer et al. 2004), which adds Ba but no
significant Eu.  This may be due to contributions from progressively longer-lived AGB stars, or
arise from the metal-dependence of the weak $s$-process from massive stars.
Naturally, the CH star in our sample stands out at [Ba/Eu]=0.98, similar to the mean of 
s-process enhanced halo stars in Roederer et al. (2014).  The [Ba/Eu] ratios in our sample
are generally slightly enhanced compared to the MW halo points, the same holds for the [$\alpha$/Eu] ratios. 
This might signal that
our Eu abundances are systematically too low.  Whether these low Eu abundances are real or spurious
is important; if real, they signal a difference between our metal-poor stars and the halo.

\begin{figure}[!htb]
\begin{center}
\includegraphics[angle=0,width=1\hsize]{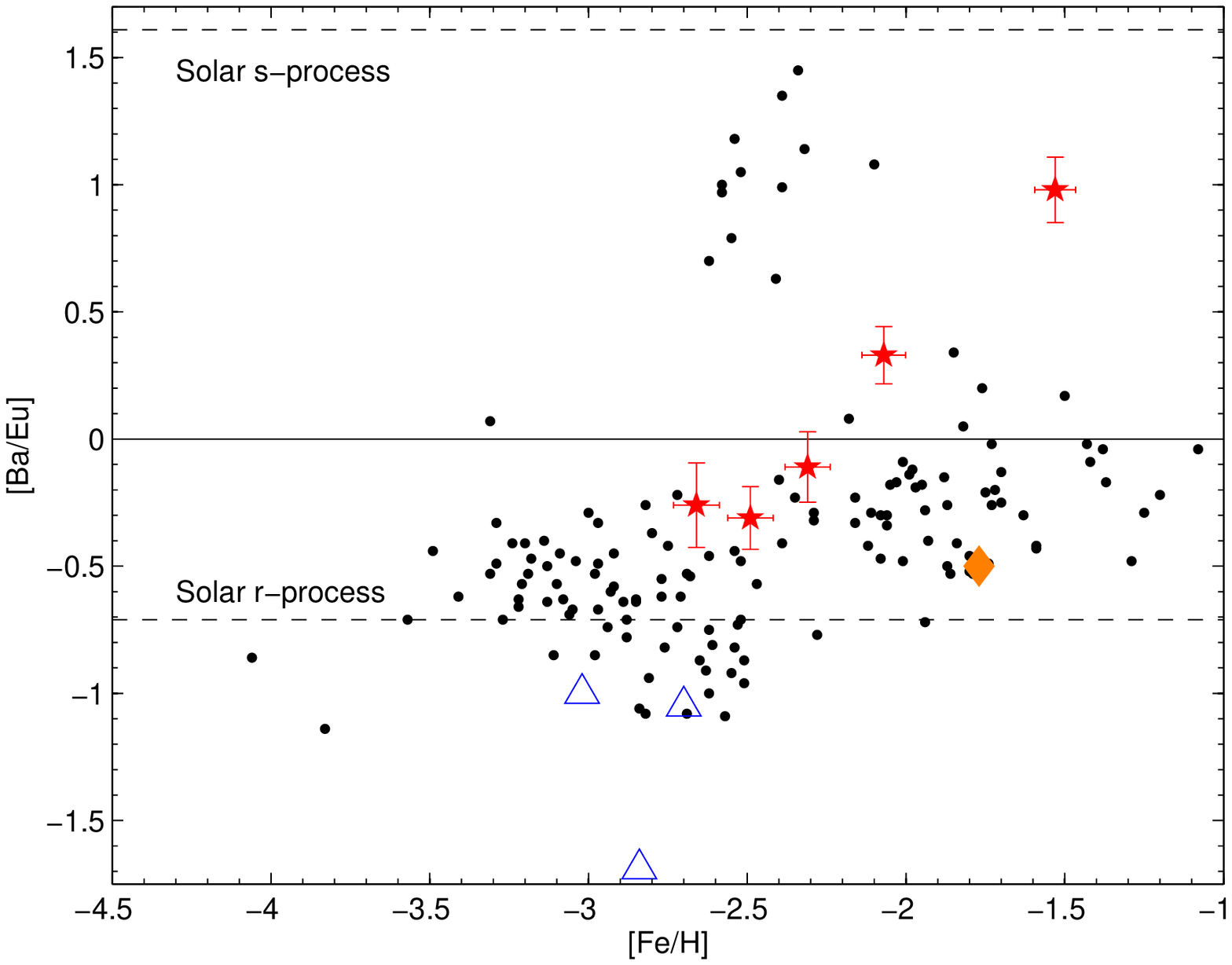}
\includegraphics[angle=0,width=1\hsize]{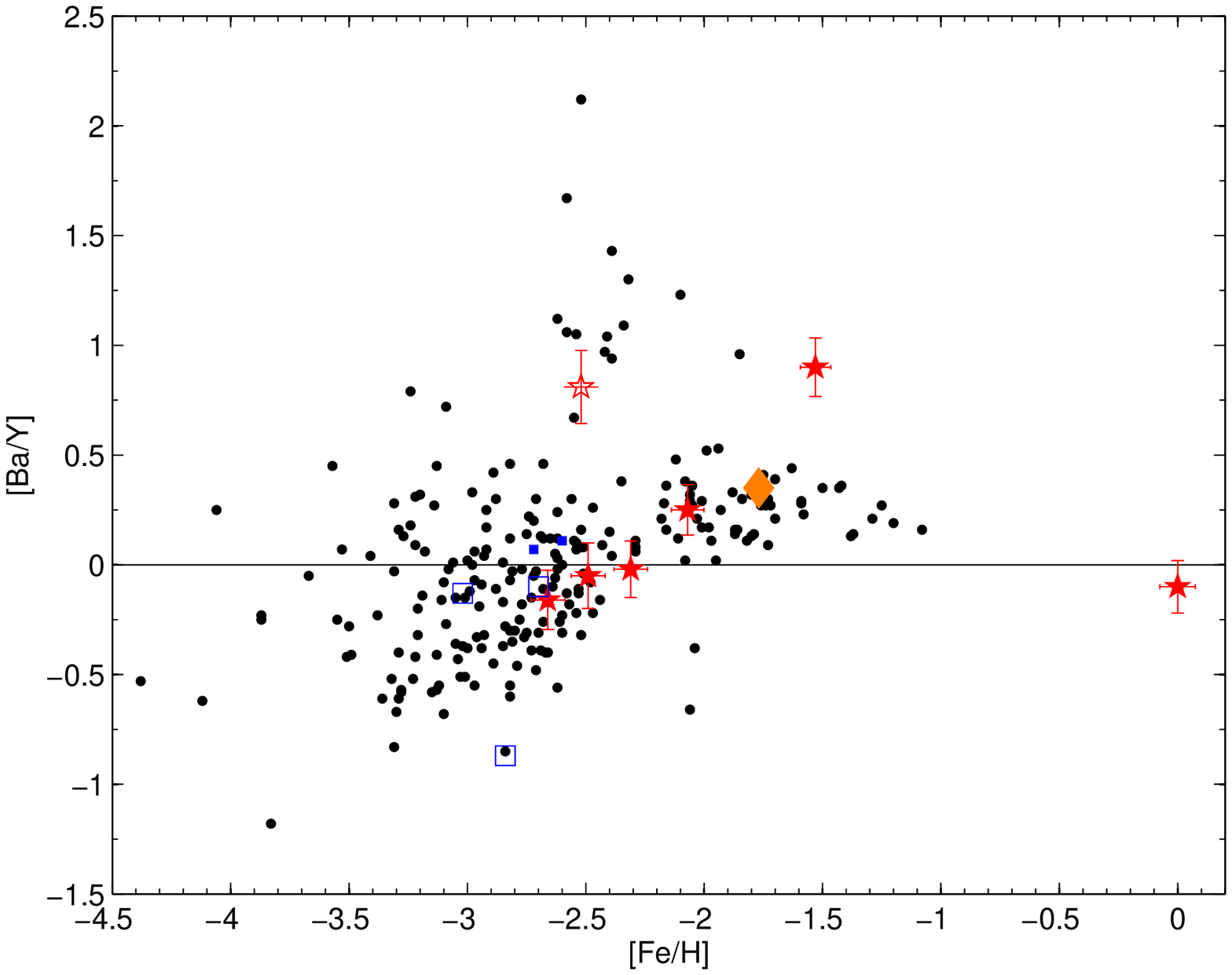}
\end{center}
\caption{{\em Top panel:} [Ba/Eu] as a tracer of the $s$/$r$-ratio. While we measure systematically higher Eu abundances in the bulge star of Johnson (2013b; yellow diamond), 
the  transition from $r$- to $s$-process dominated material is clearly seen. The dashed lines are the solar components from Burris et al. (2000) 
The {\em bottom panel} shows the [Ba/Y] as a proxy for the [$ls$/$hs$] ratio. The symbols are the same as in the previous figures.}
\end{figure}

Given the relatively high degree of homogeneity of most of the sample stars (1$\sigma$ spread of 0.25--0.50 dex over almost 0.7 dex in [Fe/H]) we show in Fig.~7 the average 
of the neutron-capture abundances of the ``regular'' stars, i.e., excluding the CH-, CEMP-$s$-, and the metal-rich stars in our sample. 
\begin{figure}[!htb]
\begin{center}
\includegraphics[angle=0,width=1\hsize]{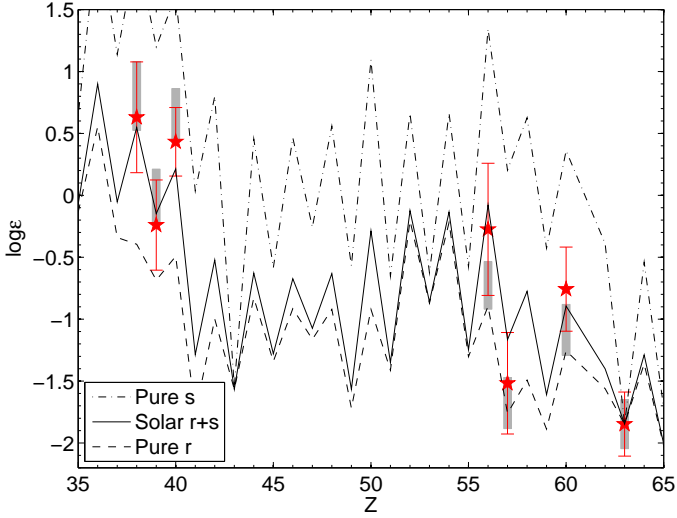}
\end{center}
\caption{Mean neutron capture element abundances to the exclusion of the CH star, the CEMP-$s$ star, and the metal-rich giant. 
The errorbars represent the star-to-star scatter. Also shown are the  Solar $r$- and $s$-process contributions from Simmerer et al. (2004). 
Grey-shaded areas indicate the 1$\sigma$ abundance ranges of the weak-$r$-process star HD 122563 from Honda et al. (2006).
All curves were normalised to Eu.}
\end{figure}
Thus the heavy element distribution in our sample metal-poor bulge candidates shows strong resemblance to the Solar-scaled 
$r+s$ distribution, with the exception of La and possibly Y, which show a preponderance of $r$-processed material.  
This is also bolstered by the comparison with the metal-poor standard star HD 122563, the neutron-capture elements of which 
are dominated by the weak $r$-process (see grey-shaded areas in Fig.~7). 
\section{Notes on individual, peculiar stars}
Here we describe the two stars with abundance peculiarities that stand out from the regular halo stars: 10464 with strong $s$-process enhancements, 
and the CEMP-$s$ star 27793.
\subsection{The CH star 10464}
The most commonly used Ba lines are very strong in this star, with EWs in excess of 250 m\AA, and are thus not on the linear part of the curve of growth. 
While they still give consistent results (at a 1$\sigma$ deviation of 0.17 dex for all 6 lines), we adopt here the abundance based on 
only the two ``weakest'' lines at 4130\AA~(EW of 118 m\AA) and 5853\AA~(176 m\AA), yielding a [Ba/Fe] ratio of +1.35 dex. 
Barium stars were first identified by  Bidelman \& Keenan (1951)
and show typical  overabundances in $s$-process elements of  0.6 to $\sim$2 dex (e.g., Allen \& Barbuy 2006). These objects also show strong CH and CN features
and, coupled with the low metallicity of star 10464 of $-1.53$ dex, this object rather qualifies 
as a Population II  CH star (Keenan 1942). 
The carbon abundance in 10464 is only mildly enhanced to [C/Fe]=0.41 dex; furthermore, the absence
of the C$_2$ band heads at 5635 and 5165\AA\ indicate that the C/O ratio is less than unity, and 
that this star is not a C-star.
We note that the stellar parameters of 10464 are very similar to the high-velocity halo CH star CD-62$\degr$1346 (Pereira et al. 2012). 

The classical explanation for strong $s$-process enhancements in stars that are not evolved
enough to have produced these elements, is mass-transfer from an AGB companion, 
the remnant of which could still be present in a binary system in the 
form of a white dwarf. 
We could not detect any evidence for radial velocity variations over the 24 hours of observations, which is, however,  clearly too short a time baseline, given 
the typical periods of Ba- and CH stars well in excess of 100 days (e.g., McClure \& Woodsworth 1990; Karakas et al. 2000). 
Likewise, no information on a possible companion is seen in the spectra or the cross-correlation function of this star, which would reveal its binarity.

The comprehensive abundance distribution enables, in principle, constraints on the AGB polluter's progenitor mass (e.g., Busso et al. 2001, Husti et al. 2009). 
We attempted a comparison with the f.r.u.i.t.y. data base (Cristallo et al. 2011), where we chose models representing the metallicity of 10464. 
An important tracer for the progenitor mass is the [Rb/Zr] ratio, which, at 0.99$\pm$0.22 dex, is very high in this star and which indicates pollution 
by  $\sim$4 M$_{\odot}$ 
AGB stars (e.g., McWiliam et al. 2013).
The detailed comparison in Fig.~8, however,  reflects the complexity of further parameters governing the AGB nucleosynthesis, such as treatment of the $^{13}$C pocket or 
the dilution factors. In particular, we find a high [$hs$/$ls$] = [$<$Ba,La,Nd$>$/Fe]$-$[$<$Rb,Sr,Y,Zr$>$/Fe] ratio of 0.41 dex, and 
relatively low abundances of La and Y.
While we do not expect any alterations of the surface abundances during the evolution towards the HB that 10464 could have experienced,  
 small enhancements of elements with overabundances of a few tenths
of a dex  could be caused by dilution of the AGB products with the pre-existing material
\begin{figure}[!htb]
\begin{center}
\includegraphics[angle=0,width=1\hsize]{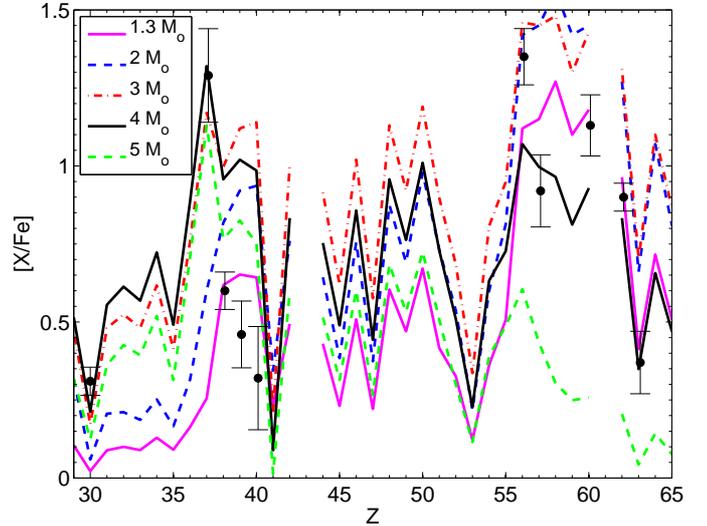}
\end{center}
\caption{Heavy element ratios in the CH star 10464
in comparison with low-metallicity ($Z$=0.0003) AGB models by Cristallo et al. (2011) for different progenitor masses.} 
\end{figure}

Lebzelter et al. (2013) detected the first two barium stars known in the bulge, from which they estimate  an occurrence rate of  $\sim$1\%, a value compatible with the
frequency found in the Galactic disks.
Johnson et al. (2012) identify a few stars with increased heavy element abundances, however, their most extreme star is consistent with dominant $r$-process
nucleosynthesis (Johnson et al. 2013). 
Thus star 10464 would establish the first known CH star in the Galactic bulge.
\subsection{The CEMP-s star 27793}
Similar to the CH star 10464, just discussed, 
the  most likely origin of the carbon- and $s$-process overabundances in 27793, a CEMP star, is
mass transfer from the envelope of an AGB binary
companion (Preston \& Sneden 2001; Lucatello et al. 2005); in fact the barium-, CH-, and CEMP-$s$ stars are often identified as the same subclasses of  
mass-transfer stars that differ only by their metallicity (e.g., Masseron et al. 2010). 
Neither in star 27793 could we detect any radial velocity variations, although the observations covered only 1.8 hours, which is far shorter
than typical orbital periods (e.g., Preston \& Sneden 2001).
In the Galactic halo, 80\% of the CEMP stars  show strong enrichments in the $s$-process elements (Aoki et al. 2007), although the numbers
differ when separated into inner and outer halo (Carollo et al. 2014). 
Norris et al. (2013) determined that the metallicity distribution of the CEMP-$s$ stars peaks at $-2.5$ dex, which is identical to 
the value we found for the bulge-candidate 27793. 

We can ask whether the level of C-enhancement is typical of a star of this metallicity. 
Spite et al. (2013) note that at metallicities above $-$3 dex,  the carbon abundance of CEMP stars is almost constant, irrespective of the their sub-classification, 
 and close to A(C)=8.25 (see also Masseron et al. 2010). At A(C)=7.33 dex, 27793 lies on the low side of the enhancement, but unmistakably qualifies as a CEMP star.  This spectrum of 27793 shows
very strong C$_2$ band heads at 5165 and 5635\AA , clearly demonstrating that C/O$>$1 and confirming its C-star 
status.

The Eu line at 4129\AA~is too deeply embedded in C-bands to be of any use. However, upper limits for the weaker $r$-process lines at 
3819,4205,4522,6645\AA~(Eu, though blended) and 5169\AA~(Dy) indicate that this star is not enhanced in material processed through $r$-process nucleosynthesis.
Finally, we note that, given its favourable location, this would be the first CEMP star detected in the bulge (but see Sect.~6.2).
\section{Membership with the bulge}
\subsection{Colour-magnitude diagrams}
Fig.~9 shows the location of our targets on the 2MASS CMD of the immediate, targeted bulge region.
Two red clumps can be clearly seen, at K=12.6 and 13.3 mag and thus with a magnitude separation compatible with 
that reported for low latitudes by McWilliam \& Zoccali (2010).  
As per our preselection criteria 
our target stars stand out as very blue objects, already hinting at their metal-poor nature against the contrasting metal-rich CMD of the main bulge population.
Likewise, the two  bluest stars in the sample are consistent  with a characterisation as metal-poor  RHB stars, 
although their very blue colours could also hint at them not being part of the bulge population. 
\begin{figure}[!htb]
\begin{center}
\includegraphics[angle=0,width=0.8\hsize]{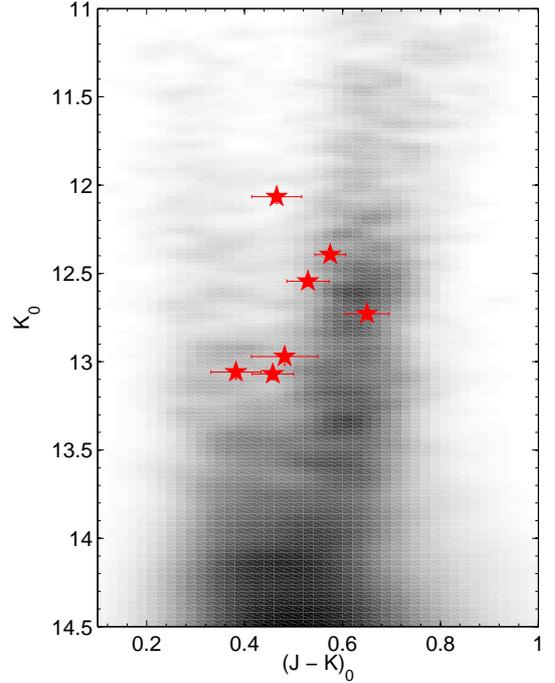}
\end{center}
\caption{Hess diagram from the 2MASS. Our targets are highlighted as error bars.}
\end{figure}
\subsection{Spatial distribution}
With our spectroscopic gravities at hand, we were able to derive distances to the stars, using the refined metallicity values that, in turn, 
allowed us to get a better estimate of the stellar masses by comparison to old, metal-poor isochrones (Dotter et al. 2008). 
Errors on the photometry and gravity were propagated into the distance uncertainties, which result on the order of 25\%.
One possible source of systematic error is our use of ionisation equilibrium to obtain the spectroscopic gravities. 
This thus required difference between Fe\,{\sc i} and  Fe\,{\sc ii}
 be sensitive to NLTE and the overall compositionof the stars, which, in turn,  affects the atmospheres' electron density (e.g., Koch \& McWilliam 2008).  

The resulting position of the stars in Galactic coordinates is shown in Fig.~10 and their distances are listed in Table~8. This assumes the distance from the Sun to the Galactic centre of 
R$_{\rm GC}$=8.34 kpc  (Reid et al. 2014).  
\begin{table}[!htb]
\caption{Positions and radial velocities}
 \centering          
\begin{tabular}{cccccr}     
\hline\hline        
Star & R$_{\odot}$  & X  & Y & Z & $v_{\rm rad}$\\  
     & [kpc] &  [kpc] &  [kpc] &  [kpc] &  [km\,s$^{-1}$] \\  
\hline        
{} &   4.0       &   4.6      &	  0.00    &  $-$0.8     &                 \\ 
\raisebox{1.5ex}[-1.5ex]{14135}        &   $\pm$0.9      & $\pm$0.8   & $\pm$0.01   & $\pm$0.2    & \raisebox{1.5ex}[-1.5ex]{$-106.9$ } \\
{} &    9.8       &  $-$1.1    & $-$0.06	  &  $-$2.0	&      \\ 
\raisebox{1.5ex}[-1.5ex]{13971}	&   $\pm$2.5   & $\pm$2.3   & $\pm$0.01   & $\pm$0.5  	& \raisebox{1.5ex}[-1.5ex]{$-$87.42} \\
{} &    15.0      &  $-$6.1    & $-$0.10	  &  $-$3.1	&        \\ 
\raisebox{1.5ex}[-1.5ex]{17221}	&   $\pm$3.2   & $\pm$4.0   & $\pm$0.03   & $\pm$0.9  	& \raisebox{1.5ex}[-1.5ex]{30.26} \\
{} &    10.4      &  $-$1.6    & $-$0.05	  &  $-$2.2	&    \\ 
\raisebox{1.5ex}[-1.5ex]{24995}	&   $\pm$2.7   & $\pm$2.6   & $\pm$0.01   & $\pm$0.6  	& \raisebox{1.5ex}[-1.5ex]{70.36} \\
{} &    9.4       &  $-$0.7    & $+$0.05	  &  $-$1.8	&    		   \\ 
\raisebox{1.5ex}[-1.5ex]{27793}	&   $\pm$2.4   & $\pm$2.2   & $\pm$0.01   & $\pm$0.4    & \raisebox{1.5ex}[-1.5ex]{$-$101.08} \\
{} &    15.6      &  $-$6.8    & $+$0.13	  &  $-$2.9	&      \\ 
\raisebox{1.5ex}[-1.5ex]{10464}	&   $\pm$5.0   & $\pm$5.1   & $\pm$0.04   & $\pm$0.9  	& \raisebox{1.5ex}[-1.5ex]{$-$27.60} \\
{} &    21.6      &  $-$12.7   & $-$0.13	  &  $-$4.1     &  	           \\   
\raisebox{1.5ex}[-1.5ex]{37860}	&   $\pm$5.0   & $\pm5.3$   & $\pm$0.03   & $\pm$1.0  	& \raisebox{1.5ex}[-1.5ex]{$-$118.22} \\
	 \hline
 \hline
 \end{tabular}
\end{table}

Due to the selection of the fields, all targets lie at low Galactic latitudes ($b\sim-11\degr$), approximately 2--4 kpc below the plane of the MW, 
and within $\sim$130 pc of the minor axis. We found, however, a large spread in heliocentric distances to the stars, from 
4.0--21.6 kpc.  
Accordingly, three objects (13971, 24995, and the CEMP-$s$ star 27793) can be associated with locations in the southernmost, far-side of the  bulge, 
whereas the most metal-rich object 14135 is situated at the near-side close to the plane. 
However, stars 17221 and the CH star 10464 lie at distances of $\sim$15 kpc, 
well beyond the bulge.

\begin{figure}[!htb]
\begin{center}
\includegraphics[angle=0,width=1\hsize]{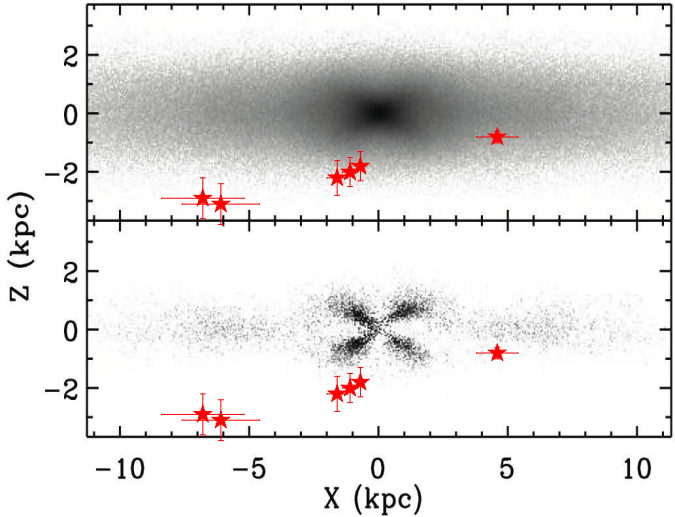}
\end{center}
\caption{Position of the target stars in Galactic coordinates, overlaid on the 
bar model (top) of Li \& Shen (2012) and the residual X-shape after subtraction of the smooth bulge component (bottom panel) from the same model.
Note that the HB star 37860 falls outside of the plotted regions.}
\end{figure}
Fig.~10 shows the stars on the model of the Galactic bulge of Li \& Shen et al. (2012), who 
identified an X-shaped component in the bar/boxy bulge models of Shen et al. (2010) 
that matched the stellar observations of McWilliam \& Zoccali (2010). The Shen et al. (2010) model was, in turn,  
tailored to reproduce the kinematic data of the Bulge Radial Velocity Assay (Howard et al. 2008; see also Wegg et al. 2015, their Fig.~1).

Li \& Shen (2010)  found an end-to-end separation of the modeled X in radial and vertical direction of 
3 and 1.8 kpc, which deems it unlikely that {\em all} our targets would be part of the X-shaped structure. Furthermore, only 7\% of the light is contained in the boxy-bulge region of the X-shape.
The comparison of the derived locations of our stars with this model (Fig.~10) suggests that those four stars, if members of the bulge,  are rather part of the 
smooth bar/boxy bulge population and not of the X. 

At $b=-11\degr$ the double red clump discovered by McWilliam \& Zoccali (2010) is very faint; accordingly,  the bulge
density has dropped off to very low values by this latitude.  
Moreover, more bulge stars are expected 
to be seen in the foreground X than the background X, due to the lower latitude (and higher 
density) of the foreground population, while we observe the opposite in that 
our stars peak behind the Galactic centre.
This distribution can be explained by  a centrally concentrated population
in that the sample volume increases with distance, while the underlying density
still has not dropped to significantly low values. 
In this line of reasoning, most of our stars were more likely to be halo stars, safe for the obvious 
member 14135, that can be associated with the bulge based on its high metallicity and abundance imprints.  

{ 
Finally, we can characterise 
the underlying radial profile of our targets, under the assumption that they are
all drawn from the same population, which  we assume to be the inner halo in the following. 
Excluding star 14135, the cumulative radial distance distribution of the remaining six stars 
would be best described by a power-law with a power-law index of about $-2.23$. 
This is indeed similar to the  decline of the stellar halo profile with power between $-2$ and $-4$ 
that is found from other tracers out to large radii (e.g., Morrison et al. 2000; Yanny et al. 2000; Sesar et al. 2011; Faccioli et al. 2014); 
note that Blanco \& Terndrup (1989) already reported on a steep density profile $\propto$R$_{\rm GC}^{-3.65}$ for the inner spheroid.  
We caution, however, that this 
is  clearly hampered by low-number statistics, in particular at Galactocentric distances outside of 5 kpc, 
and the largely uncertain selection function of our targets from an underlying, true, inner halo population.} 
\subsubsection{Star 37860 and Sagittarius}
The RHB star 37860 lies at 21.6 kpc to the observer, which rules out a physical connection with the bulge.
Given this distance, its position within the Galaxy, and its low radial velocity of $-118$ km\,s$^{-1}$ it is plausible that 
37860 is a member of the tidally disrupted Sagittarius (Sgr) dwarf galaxy (Ibata et al. 1994). 
While Sgr's main body shows very different velocities of $\sim$140 km\,s$^{-1}$ (Ibata et al. 1994), 
the simulations of Fellhauer et al. (2006) predict parts of the old trailing arm to pass near the region and distance where our star 37860 is located, at the 
large negative radial velocity we indeed observed. The best match in this parameter space is with the component of the Sgr stream that  was 
stripped from the dwarf galaxy in the simulations more than 7.4 Gyr ago. 

Stars as metal poor as 37860 ([Fe/H]=$-2.07$) are known to exist in various components associated with the Sgr system, including field RR Lyrae (Vivas et al. 2005)
and globular clusters stripped from the disrupted dwarf (e.g., Mottini \& Wallerstein 2008; Law \& Majewski 2010), each down to $-2.3$ dex. 
If 37860 was indeed a member of the metal-poor Sgr population, its regular abundance pattern points towards a chemical evolution of those parts that were stripped early on in 
the accretion process that was typical for similarly old, metal-poor environments (e.g., Roederer et al. 2014; Koch \& Rich 2014), contrasting
the complex evolution of Sgr's compact and intact main body (McWilliam et al. 2013), at higher [Fe/H]. 
\subsection{Kinematics}
In order to further investigate the nature of our sample stars, we extracted their proper motions from two sources: the Southern Proper Motion Catalog 4 (SPM4; Girard et al. 2011) 
and The Fourth US Naval Observatory CCD Astrograph Catalog (UCAC4; Zacharias et al. 2013). 
As Fig.~11 implies, the proper motions from both surveys are significantly different for more than half of our sample, which hampers a reliable assessment of the stars' dynamical history and in the following, we
only briefly comment on a few points { (see also the discussion in Casey \& Schlaufman 2015 and their Table~2)}. 
\begin{figure}[!htb]
\begin{center}
\includegraphics[angle=0,width=1\hsize]{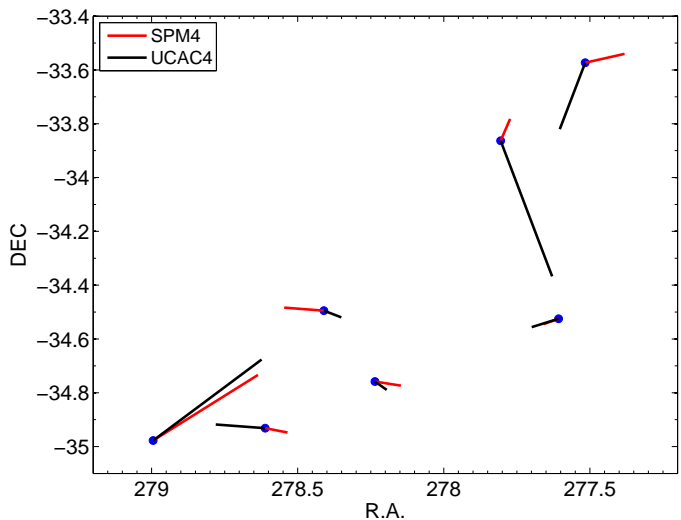}
\end{center}
\caption{Comparison of proper motions from two different catalogues. The direction of the motion is indicated by lines protruding at the positions of each object
with lengths proportional to the absolute values of the proper motion.}
\end{figure}

Using the stellar kinematics, 
we integrated their orbits 12 Gyr backwards in the Galactic potential of  Dehnen \& Binney (1998), which contains contributions from the halo and disk. 
Here, the bulge is assumed to follow a spherical density distribution. 
We note that only in two cases, realistic (``bound'') orbits could be obtained 
 in the sense that the orbital period was below a Hubble time. All others are 
on radial passages and periods or apocentres cannot be further quantified.
Taken at face value, this would imply that the majority of our stars would be on high-velocity, radial orbits that would make them pass through the halo and bulge regions as mere interlopers 
(e.g., Kunder et al. 2015). But we emphasise again that these interpretations are subject to the largely uncertain proper motions and need consolidation from future, improved kinematic data, e.g., from the Gaia
astrometric mission.
\section{Summary and conclusions}
The main scope of this work had been to identify and analyse metal-poor star candidates in the Galactic bulge so as to trace the earliest enrichment phases of 
one of the oldest components of the MW.  
Our main findings from an  analysis of seven stars located towards the  bulge are:
\begin{itemize}
\item Out of the five red giants and two horizontal branch stars of the sample, five are in the very metal-poor regime, at $-2.7<$[Fe/H]$<-2.0$. 
The most metal-rich star is a typical bulge member at Solar metallicity with Solar abundance ratios in  all elements studied.
\item We do not see any evidence for chemical enrichment by massive Population III stars: 
all stars have approximately Solar [Sc/Fe] ratios, in contrast to nucleosynthetic models of those massive stars,  
which are expected to have occurred in large numbers in the central regions of the MW.
\item The  metal-poor sample  is chemically 'normal' in the remainder of the determined elements:  they show chemical abundance ratios that are typical of metal-poor halo stars, or
of their respective class such as CEMP-$s$ and CH-stars. If these stars are indeed members of the bulge, this indicates that the central regions of our Galaxy 
chemically resemble the inner halo. 
\item The two stars that show strong enhancements in carbon and/or $s$-process elements, provide, to our knowledge,  the first detection of a CEMP and CH star towards the Galactic bulge. 
\item Based on its position on the sky, its distance, and radial velocity, one star is consistent with the old, trailing arm of the disrupted Sagittarius dwarf galaxy. 
Its low metallicity gives important insight into the early enrichment and disruption history of this stellar system.
\item From our spectroscopic distances we cannot refute that three of our target stars are part of the far-side of the X-shaped bulge.
\item Overall, the chemical abundances and distances of our stars suggest that an association with the inner halo is likely.
Given the broad mix of stars we discovered, it is possible that we see halo stars on their orbit through the central regions of the Galaxy.  
Thus any detection of genuine metal-poor ``bulge'' stars needs to be taken with caution and accurate kinematic information is imperative to 
consolidate their membership. 
\end{itemize}
Furthermore, our results  show that, even with a careful selection procedure, no unambiguous
detection of one single, homogeneous  ``metal-poor bulge''   could be obtained.
Further selections using infrared photometry are currently being explored and offer promising results (e.g., Schlaufman \& Casey 2015).
\begin{acknowledgements}
AK acknowledges the Deutsche Forschungsgemeinschaft for funding from  Emmy-Noether grant  Ko 4161/1. 
We thank the anonymous referee for a helpful report.
This research made use of atomic  data from the INSPECT database, version 1.0 (www.inspect-stars.com). 
 \end{acknowledgements}
\end{document}